\pdfoutput=1
\documentclass{aa}  

\usepackage{graphicx}
\usepackage{upgreek}
\usepackage{txfonts}
\usepackage[breaklinks=true]{hyperref}
\begin{document}

   \title{Faint polarised sources in the Lockman Hole field at 1.4\,GHz}

%   \subtitle{I. Overviewing the $\kappa$-mechanism}

   \author{A. Berger\inst{1}
          \and
          B. Adebahr\inst{1}
          \and
          N. Herrera Ruiz\inst{1}
          \and
          A. H. Wright \inst{2}
          \and
          I. Prandoni\inst{3}
          \and R.-J. Dettmar\inst{1}}

   \institute{Ruhr University Bochum, Faculty of Physics and Astronomy, Astronomical Institute (AIRUB), Universitätsstrasse 150, 44780 Bochum, Germany
   \and Ruhr University Bochum, Faculty of Physics and Astronomy, Astronomical Institute (AIRUB), German Centre for Cosmological Lensing, Universitätsstrasse 150, 44780 Bochum, Germany
   \and INAF - Instituto di Radioastronomia, via Gobletti 101, I-40126 Bologna, Italy}

   \date{}

% \abstract{}{}{}{}{} 
% 5 {} token are mandatory
 
  \abstract
   {In the context of structure formation and galaxy evolution, the contribution of magnetic fields is not well understood. Feedback processes originating from AGN activity and star-formation can be actively influenced by magnetic fields,  depending on their strength and morphology. One of the best tracers of magnetic fields is polarised radio emission.   Tracing this emission over a broad redshift range therefore allows investigation of these fields and their evolution.}
   {We aim to study the nature of the faint, polarised radio source population whose source composition and redshift dependence contain information about the strength, morphology, and evolution of magnetic fields over cosmic timescales.}
   {We use a 15 pointing radio continuum L-band mosaic of the Lockman Hole, observed in full polarisation, generated from archival data of the Westerbork Synthesis Radio Telescope (WSRT). The data were analysed using the RM-Synthesis technique. We achieved a noise of 7\,$\mu$Jy/beam in polarised intensity, with a resolution of 15\arcsec. Using infrared and optical images and source catalogues, we were able to cross-identify and determine redshifts for one third of our detected polarised sources.}
   {We detected 150 polarised sources, most of which are weakly polarised with a mean fractional polarisation of 5.4\,\%. No source was found with a fractional polarisation higher than 21\,\%. With a total area of $6.5\,$deg$^2$ and a detection threshold of 6.25\,$\sigma$ we find 23 polarised sources per deg$^2$. 
   Based on our multi wavelength analysis, we find that our sample consists of AGN only.%All detected polarised sources were classified as AGN. 
   We find a discrepancy between archival number counts and those present in our data, which we attribute to sample variance (i.e. large scale structures). Considering the absolute radio luminosty, to distinguish weak and strong sources, we find a general trend of increased probability to detect weak sources at low redshift and strong sources at high redshift. We attribute this trend to a selection bias. Further, we find an anti-correlation between fractional polarisation and redshift for our strong sources sample at $z\geq0.6$.}
   {A decrease in the fractional polarisation of strong sources with increasing redshift cannot be explained by a constant magnetic field and electron density over cosmic scales, however the changing properties of cluster environments over the cosmic time may play an important role. Disentangling these two effects requires deeper and wider polarisation observations, and better models of the morphology and strength of cosmic magnetic fields.}

   \keywords{Magnetic fields -- Polarization -- Cosmology: observations --  Galaxies: active -- Radio continuum: general}

   \maketitle
%
%-------------------------------------------------------------------

\section{Introduction}

Feedback processes within starforming galaxies and active galactic nuclei (AGN) are known to be the main drivers of the enrichment of the intergalactic medium. While star-forming galaxies are more numerous than AGN by two orders of magnitude, AGN are much more energetic. It is therefore a matter of debate, which of these two object classes dominates the feedback process.

It is known that AGN and star-formation activity correlate up to redshifts of $z\sim 1.5$ \citep{1998MNRAS.293L..49B,1998ASSL..226..129W}, but discrepancies exist at higher redshifts \citep{2005A&A...434..133W}. However many analyses of sources at higher redshifts are biased towards more massive and active sources \citep{2004PThPS.155..202T,2012A&A...537A..16F}, such as starburst objects or those in the Fanaroff-Riley (FR) II category \citep{Fanaroff1974}. As a result, the contribution of the more numerous, radio-quiet (and therefore faint) sources remains unknown. A detailed understanding of the faint radio source population, and the underlying physical processes dominating feedback, is therefore indispensable when attempting to explain galaxy formation and evolution overall.

Understanding the influence of magnetic fields on large scales, and how they evolve into the regular magnetic field structures we see in the nearby Universe today, is an open question in current observational and theoretical astronomy. While we can directly observe the morphology (and estimate the strength) of magnetic fields in nearby objects \citep{2015A&ARv..24....4B,2019ARA&A..57..467B}, the necessary information for evolutionary studies over cosmic timescales can only be provided statistically using samples of sources that span cosmic timescales. 

One of the best tracers of magnetic field strength and morphology is total and polarised radio synchrotron emission: we can examine the integrated quantities of the magnetic field via the total power emission, and analyse the field's degree of ordering and geometry using the polarised emission. 
As such, the faint polarised radio sky therefore provides a direct window into the evolution of the magnetic field over cosmic timescales  . Using statistically significant samples of polarised radio sources allows us to simultaneously trace their characteristics as a population \citep{2016ApJ...829....5L,2014ApJS..212...15F}, observe possible changes they undergo during their evolution, and explore the structure of the intervening cosmic magnetic field \citep{2018MNRAS.475.1736V}.

Most studies of polarised radio sources in the past were based on single objects or small targeted samples. Upcoming wide-field radio surveys, observing both total intensity and polarisation, are now providing statistical constraints on the polarisation properties of faint radio-selected AGN and starforming objects.  

Deep field polarisation studies at 1.4\,GHz, down to detection thresholds of around 50\,$\mu$Jy/beam and over areas of several square degrees, have been conducted within several projects \citep{Taylor2007,Grant2010,Subrahmanyan2010,Hales2014b}. \cite{Rudnick2014} made even deeper observations down to a noise level of $\sim3\,\upmu$Jy/beam but for a smaller survey area of approximately $0.3\,{\rm deg}^2$.

Total power source counts show that, below flux densities of $0.5-1\,$mJy, starforming galaxies become progressively dominant over AGN, which constitute the majority of the sources at higher flux densities \citep{Hopkins2003,2008A&A...477..459M}. These differences in structure, as well as in the source nature, likely introduce differences in the polarisation properties of the faint source population compared to their bright population counterparts  \citep{Stil2014}. In contrast to this the faint, polarised sky (down to the $\upmu$Jy level) is still dominated by AGN \citep{Hales2014b}. In addition, the polarisation properties of AGN have been found to differ depending on their morphology \citep{Conway1977} and spectral properties \citep{Mesa2002,Tucci2004}.

Studies by \citet{Mesa2002}, \citet{Tucci2004}, \citet{Taylor2007} and \citet{Grant2010} observed an anti-correlation between the degree of linear polarisation and the total intensity of the faint extra galactic radio source population at 1.4\,GHz. It was supposed by \cite{Hales2014b} that this correlation is not caused by physical properties of the sources but rather originates from incompleteness affecting the faintest sources in the sample. In agreement with this, \cite{Stil2014} observed a much more gradual trend when stacking polarised sources in the NRAO VLA Sky Survey (NVSS) \citep{Condon1998}.

The Lockman Hole \citep{Lockman1986} is one of the best studied regions of the sky over a multitude of wavelength regimes, which leads to an extensive multi-band coverage of the field. Here we briefly describe the observations we used for this publication. A complete description of all available data in the field is available in \citet{LH_Total}.
The Lockman Hole has a low infrared (IR) background of about $0.38\,{\rm MJy\,sr}^{-1}$ at 100\,$\upmu$m \citep{Lonsdale2003}, which makes it well suited for IR observations. 
Thus about $12\,\text{deg}^2$ of the field were observed with the Spitzer Space Telescope \citep{Werner2004} as part of the Spitzer Wide-area Infrared Extragalactic survey \citep[SWIRE;][]{SWIRE} at 3.6, 4.5, 5.8, 8.0, 24, 70 and 160\,$\mu$m.
The Lockman Hole was also observed in the far-infrared as part of the {\it Herschel} Multi-tiered Extragalactic Survey \citep{Oliver2012}, in the ultraviolet regime by the Galaxy Evolution Explorer (GALEX) GR6Plus7 \citep{Martin2005} and as part of the Sloan Digital Sky Survey (SDSS) DR7 \citep{Abolfathi2018}.
The Lockman Hole is also part of a deep optical weak lensing analysis by \cite{Hendrik}, who used data from the Canada-France-Hawaii Telescope (CFHT) in five optical bands ($ugriz$), and measured photometric redshifts. 
%Additionaly to the SWIRE wavelength $16\,\text{deg}^2$ of the field were observed in longer far-infrared (FIR) wavelength by the Herschel Space Observatory using the Photconductor Array Came

A variety of radio surveys cover limited areas ($<1\,{\rm deg}^2$) within the Lockman Hole region. Recently, wider radio coverage has been obtained with the Low-Frequency Array (LOFAR) at 150\,MHz \citep{Mahony2016} and with the Westerbork Synthesis Radio Telescope (WSRT) at 1.4\,GHz. The WSRT mosaic covers $\sim6.6\,{\rm deg}^2$ down to 11\,$\upmu$Jy/beam in root-mean-squared (RMS) deviation \citep{LH_Total}. Both datasets were, however, only analysed in total power radio continuum; this paper aims to study the polarisation properties of the field at 1.4\,GHz. Deep, wide-area radio surveys represent excellent probes of magnetic fields over cosmologically  significant redshift intervals. 

The paper is organised as follows. In section 2 we present our data and data reduction strategy to obtain a mosaic in polarised intensity. In section 3 we describe our source identification and catalogue creation. Section 4 covers the cross identification procedures with other source catalogues at other wavelength. This information is used in section 5 for the classification of the polarised sources. In section 6 we present our results. A part of our results are analysed with respect to the cosmic evolution of magnetic fields in section 7.
Section 8 gives a discussion of these and our summary is presented in section 9. 
%--------------------------------------------------------------------
\section{Data}
\label{sec:Data}

This study is based on the WSRT observations of the Lockman Hole field at 1.4\,GHz \citep{LH_Total}. The data consist of 16 individual pointings, each observed for a full synthesis of 12\,hrs between December 2006 and January 2007. Individual pointing centres are given in Table \ref{T:Pointings}. The centre of the mosaic was chosen to be at RA = 10:53:16.6; Dec. = +58:01:15 (J2000). 

Each dataset was recorded over a full bandwidth of 160\,MHz organised in eight 20\,MHz sub-bands with 64 channels each. The channel width is 312\,kHz, the central subband frequencies are 1451, 1422, 1411, 1393, 1371, 1351, 1331 and 1311\,MHz. The data were recorded in all four linear correlations (XX, XY, YX, YY) and thus contains full polarisation information. The total power calibration and imaging was already performed by \cite{LH_Total}.

The standard flux calibrator source 3C48 was observed for 900\,s before all 16 target fields. The polarisation calibrator 3C138 was also observed for 900\,s before the target fields, but was missing for Pointing 12, which we therefore excluded from the data reduction; inclusion of this data would have resulted in inconsistent polarisation calibration within parts of the dataset.

\subsection{Data reduction}
For the data reduction we used a combination of the Astronomical Image Processing System \cite[AIPS;][]{AIPS}, the Common Astronomy Software Application package \citep[CASA;][]{CASA} and the Multichannel Image Reconstruction Image Analysis and Display \citep[MIRIAD;][]{MIRIAD} software package.
First AIPS was used to apply the system temperatures, after which we converted the data into the CASA MS-format to inspect the data for radio frequency interference (RFI). 
We used AOFlagger \citep{Offringa2010} to perform an automatic flagging of RFI within our dataset. 
To allow easier identification of RFI we applied a preliminary bandpass calibration, thereby mitigating the rapid rolloff effect of the receiver response curve at the edges of the frequency bands. This bandpass, however, was not used for any further calibration steps.
All data were additionally inspected by eye to flag remaining RFI using RFI GUI (the graphical front-end of AOFlagger).
Afterwards, the final bandpass calibration was applied, as well as the gain and polarisation leakage calibration, using the unpolarised calibrator source 3C48.
The calibrator 3C138 was used to calibrate the polarisation angle. 
%Since neither the calibrator source 3C138 nor another well known polarisation calibrator was observed for one pointing this pointing got excluded from further analysis. 
Models for 3C48 and 3C138 were taken from \citet{Perley2017}. The cross-calibration was then performed on a per-channel basis following the procedure described in \cite{2013A&A...555A..23A}, to minimise polarisation leakage.
After the cross calibration in CASA, we imported the target data into MIRIAD for imaging and full polarimetric self-calibration using the MIRIAD task GPSCAL.
To avoid including artificial sources in the self-calibration process, we created masks for every pointing in total power and used them for cleaning all four Stokes parameters. The self-calibration and imaging process was performed on each subband for each pointing individually.
We consecutively decreased the solution interval from 20 minutes (for the first self-calibration cycle) to 30\,s (for the last). We excluded short baselines in the first self-calibration cycles and extended the (u,v)-range to include all baselines for the latter ones.
Lastly, we used phase-only solutions for the first iterations and, in the final step, included both amplitude and phase calibration for data where enough signal-to-noise was available. 

Using a joint deconvolution approach we generated cleaned image cubes of all 15 pointings in Stokes Q and U. For each subband the data of eight adjacent channels was imaged together resulting in an averaged channel width of 2.5\,MHz each. Averaging the data to a coarser frequency resolution before imaging has the advantage that the individual Stokes Q and U images can be cleaned to greater depth, and therefore artefacts from sidelobes can be minimised. The resulting 128 images (64 for each of the two Stokes parameters) covering the whole bandwidth of the observation were cleaned and independently primary-beam-corrected using the standard primary beam correction for the WSRT: $cos^6(c\nu r)$, where $c=68$ at L-band frequencies, $\nu$ is the frequency in GHz, and $r$ the distance from the pointing centre in radians.

\subsection{Rotation measure synthesis}
To mitigate the effect of bandwidth depolarisation we used the Rotation Measure (RM)-synthesis technique \citep{Brentjens2005}. The resulting parameters for the RM-Synthesis are: $\delta\psi = 328.46\,{\text{rad}}/{\text{m}^2}$, max-scale$ = 74.46\, {\text{rad}}/{\text{m}^2}$ and $\psi_{max} = 10095.57\, {\text{rad}}/{\text{m}^2}$, where $\delta\psi$ represents the resolution in Faraday space, max-scale the maximum observable width of a polarised structure in Faraday space and $\psi_{max}$ the maximum observable Faraday depth before polarised emission becomes depolarised due to bandwidth smearing effects.
Frequency averaging only influences the maximum observable RM and the recovered polarised intensity in case of sources exceeding this limit. For this study we expect our sources to have maximum RMs of several hundred, such that we ought not to lose any relevant information due to the applied averaging.

\begin{figure*}
    \centering
    \includegraphics[width = \textwidth]{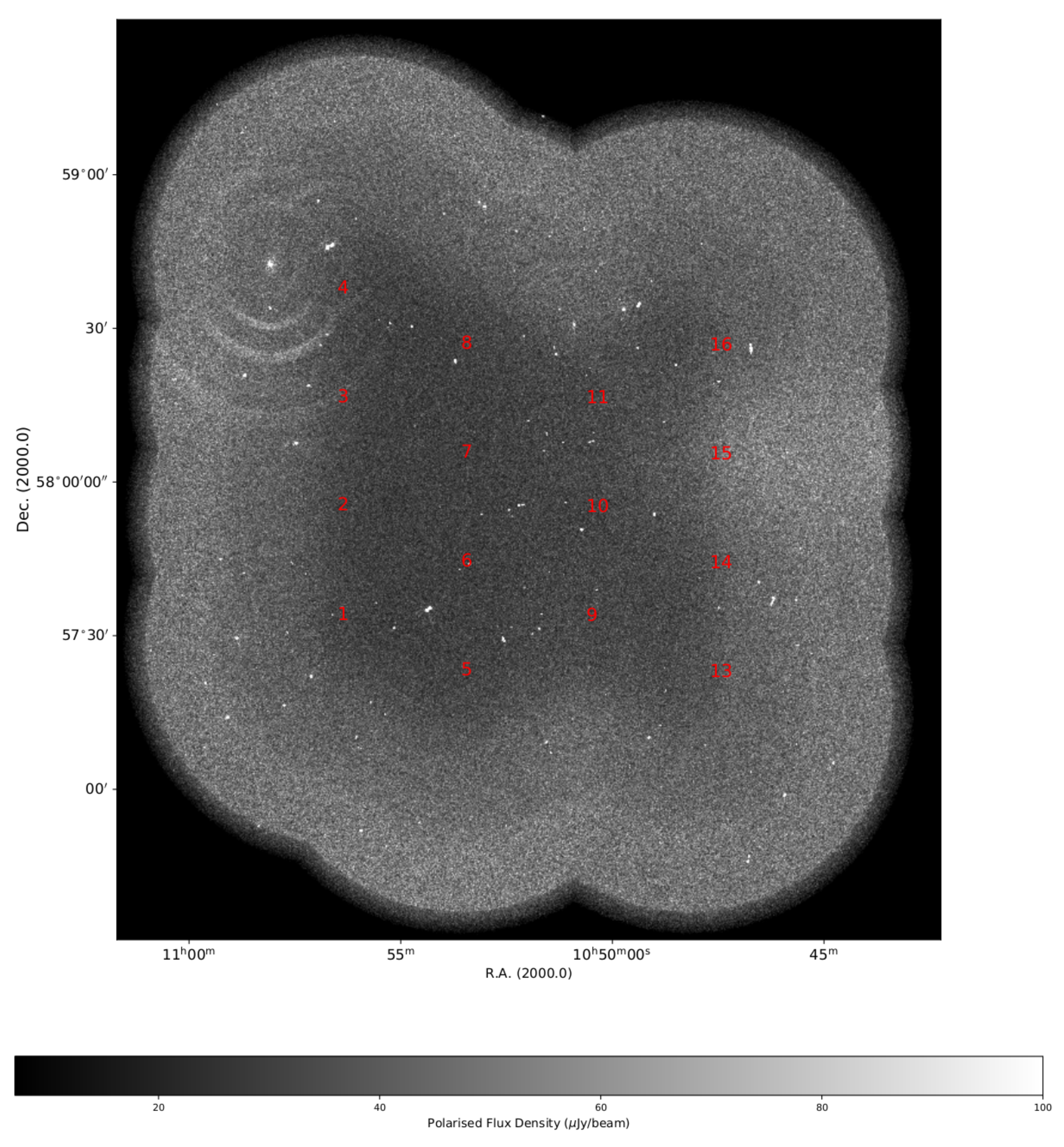}
    \caption{1.4\,GHz polarised intensity mosaic from 15 overlapping pointings observed with the WSRT. The greyscale shows the polarised flux density. The centres of the individual pointing positions are marked with the pointing numbers in red. Pointing 12 was not used in this analysis, as discussed in sec. \protect \ref{sec:Data}.}
    \label{PImap}
\end{figure*}

The Faraday cube was sampled between $-1024\,{\text{rad}}/{\text{m}^2}$ and $1024\,{\text{rad}}/{\text{m}^2}$ using a sampling interval of $4\,{\text{rad}}/{\text{m}^2}$.
To derive the polarised emission map % and the rotation measure map
from the cube, we fitted a parabola to the first peak along the Faraday axis at any values exceeding 5$\sigma$.
The polarised emission is determined by the peak value of the parabola. % while the RM is given by the position of the peak.
This was done to compensate for the limited sampling rate.
Our final polarised emission map reaches a central RMS of $7\,\mu$Jy/beam. The map is shown in figure \ref{PImap}.

%-----------------------------------------------------------------

\section{Source detection}

\subsection{pyBDSF}

We used pyBDSF \citep{pyBDSF} to create a catalogue of polarised sources. 
To account for the higher RMS close to the edges of the mosaic we derived a local RMS map by using the adaptiv\_rms\_box option with 60px boxes with a stepsize of 20 px.
%used the adaptive rms box with a 60\,px box and a step size of 20\,px. 
%To this aim we also forced the programme to calculate a 2D rms map, as well as a 2D mean map.
As a source detection threshold we used 6.25\,$\sigma$, which is comparable to a Gaussian threshold of 5.33\,$\sigma$ \citep{Hales2014b}. The threshold for the island boundary was set to 5\,$\sigma$.
This resulted in a catalogue containing 178 components. 
To acquire a measure of the complete polarised flux of each source, we summed the flux densities of the associated components (as identified by pyBDSF) which resulted in a catalogue of 154 polarised sources.
The summation was done with the measurements of pyBDSF on the polarised intensity map.
We note that following sec. \ref{sec:inst_pol} the final published catalogue contains 150 sources.

\subsection{Fractional polarisation}

To calculate the fractional polarisation values ($\Pi = \text{PI/I}$, where PI is the polarised intensity and I the total intensity) of every source we also created an image in total intensity. 
%By comparing our total intensity map to the total power image published by \cite{LH_Total} and to the Faint Images of the Radio Sky at Twenty-Centimeters \citep[FIRST;][]{FIRST} survey, we found a few sources in the image of \cite{LH_Total} as double or triple sources, which are single sources in FIRST, as well as in our total intensity map. These multiple sources appeared exclusively in the overlapping regions of two or more pointings. 
%Since all of the problematic sources are well above the detection limit of the FIRST survey and compact enough to be detected by it, we concluded that the \cite{LH_Total}-mosaic is affected by systematic data reduction or imaging errors. For consistency reasons we therefore decided to use our own total power image for any further analysis. 
Importantly, we opted to generate a bespoke total intensity map using our reduction strategy described previously \citep[rather than simply use the map presented in][]{LH_Total}, in order to ensure consistency between our polarised and total intensity maps. %Since we used a different data reduction strategy and ended up with a different beam in our polarised intensity map than \cite{LH_Total} in their total intensity map, we decided to create our own total intensity map instead of using the one of \citet{LH_Total} to be fully consistent.
To this end we assembled a mosaic for each subband and stacked the individual subband images using the MIRIAD task IMCOMB, with an inverse square weighting of the noise in the individual images. 
The resulting mosaic has a central RMS of ~30\,$\upmu$Jy/beam.
Again for the sake of consistency, we once more used pyBSDF for our source extraction. In this case we applied a detection threshold of 5\,$\sigma$ to create a catalogue of the total power components, and the threshold for the island boundary was set to 3\,$\sigma$.
To create a source catalogue we again summed the  total flux densities of the associated components, resulting in a catalogue of 1708 sources, following the same strategy as for the polarised catalog. This leads to a source density of about 262/deg$^2$, which is in good agreement with other deep field studies with comparable noise levels \citep[eg.][]{Taylor2007,Hales2014a,Subrahmanyan2010}.
Finally, we note that the aim of the image generation in total intensity was not to optimise the detection and flux measurement for all possible sources within the field, but rather to obtain an estimate of the total intensity fluxes for each of our polarised sources that is consistent with our polarised measurements. As such, we prioritise consistency over absolute signal-to-noise, and are therefore not concerned that our resulting total intensity maps have higher central RMS than the equivalent map produced by \cite{LH_Total}. %The higher noise compared to \citep{LH_Total} who used the same data 
%\cite{LH_Total} found 5997 components including 183 multiple-component sources. 
%We have to mention here that our total intensity map has a central RMS of about 30$\,\mu$Jy, while the map from \cite{LH_Total} reaches a higher sensitivity with a central RMS of 11\,$\mu$Jy, we thus detect less sources than they were able to. 
%So we found that $9\%$ of the sources within the field show polarised emission.
%The total power image published by \citep{Pradoni2018} shows some sources as double or triple sources, compared to the equivalent detection in first. 

We cross matched the component catalogue in polarised intensity with our component catalogue in total intensity, using a matching radius of 30\arcsec; that is, twice the 15\arcsec\  beam present in the polarised intensity map.
All matches were confirmed or rejected via visual inspection. 
In case of some complex polarised sources, individual parts of the sources were identified as separate objects. To mitigate this effect, in cases where these could be associated (by visual inspection) with the same total power source, the previously individual components were again manually summed together into a single combined source for further analysis.
For two of our polarised sources, no total intensity counterpart was found. The visual inspection of these sources showed that, due to reduction artefacts in the total intensity image, pyBDSF was not able to detect these sources. These sources were therefore resigned to inherit the total flux measurements of \cite{LH_Total}, and were marked as such within our catalogue.

%All total intensity sources that showed polarised emission were than again cross matched by their isl-ID to ensure that every component is considered for further analysis.
%The total power sources were than again matched with their polarised counterparts by ID.
%The fractional polarisation is just defined by
%\begin{equation}
%    p = \frac{PI}{I}
%\end{equation}
%where PI is the polarised intensity and I is the total intensity.

\subsection{Instrumental polarisation}
\label{sec:inst_pol}

We checked for the influence of artificial instrumental polarisation by imaging four overlapping pointings. This allowed us to determine the polarised intensity for the same sources with different distances and directions from their pointing centres. 
pyBDSF was again used for source identification and flux measurements.
By cross-matching the resulting catalogues of the individual pointings with each other and with our catalogue, we found that the measured polarised flux densities are consistent within their individual uncertainties.
%comparing the fractional polarisation of the sources that are in those four pointings. 
In addition, we found no sources that were detected in an individual pointing and not detected in the mosaic, nor any sources which were detected in one individual pointing and not in another (overlapping) pointing. 

Since we therefore have no evidence for additional instrumental polarisation, we concluded to use a conservative cutoff of $0.5\,\%$ in fractional polarisation for considering sources as physically polarised. 
We excluded all sources with a fractional polarisation lower than this cutoff limit. This criterion excluded 4 sources from further analysis, resulting in a remaining sample of 150 sources with 172 components, this is a polarised source density of 23/deg$^2$. The catalogue is given in the Appendix \ref{A:Catalogue}.
Most sources were found to have modest polarisation, with a mean fractional polarisation $\Pi = 5.4\,\%$. The highest degree of polarisation found was 21\,\%.
%To avoid for false polarisation defections we used a cutoff in fractional polarisation of 0.5\,\%. This criterion excluded 4 sources from further analysis, so we have a remaining sample of 150 sources.

%-----------------------------------------------------------------

\section{Cross-identification}

\begin{figure}
    \centering
    \includegraphics[width = \linewidth]{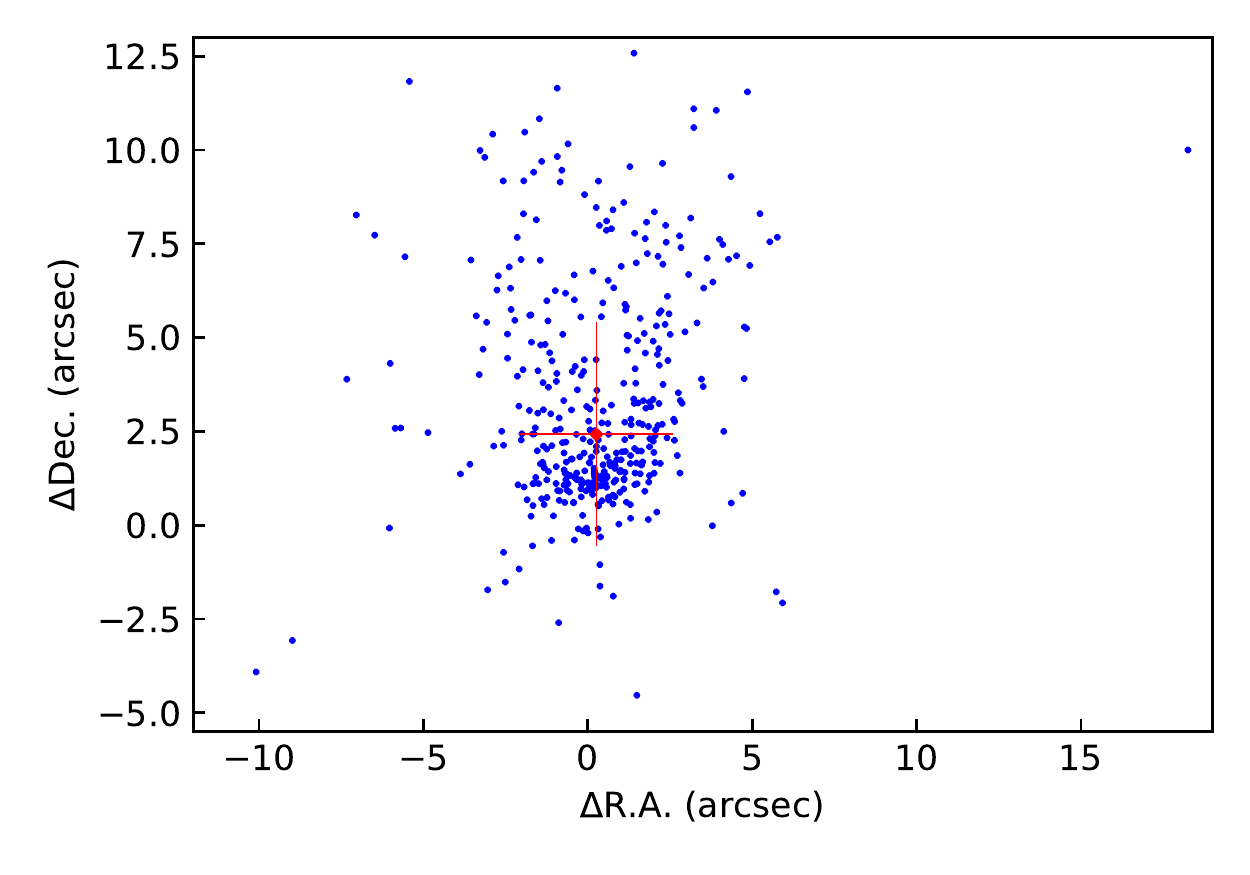}
    \caption{Differences in the source positions between our observations and the FIRST catalogue. The red cross shows the mean $\Delta \text{Ra}$ and $\Delta \text{Dec}$ of the scatter where the uncertainties are the standard deviation.}
    \label{koordinaten_offset}
\end{figure}

Using the catalogue from \cite{Helfand2015} we cross matched our total intensity sources with their FIRST counterparts to verify our astrometric calibration. To establish the calibration accuracy, we restrict this comparison to point-sources within our sample, which are selected using the S-Code ``S'' from pyBDSF. We find a median offset of $\Delta \text{RA} = 0.26\arcsec\pm 2.34\arcsec$ and $\Delta \text{DEC} = 2.42\arcsec\pm 2.99\arcsec$ of our coordinates compared to the FIRST coordinates (Fig. \ref{koordinaten_offset}). 
%The sources with larger offsets are mainly located near the edges of the mosaic. 
A general offset of the FIRST coordinates was previously noticed by \cite{Grant2010}, who found an average offset of $\Delta \text{RA} = 0.5\arcsec\pm 0.2\arcsec$ and $\Delta \text{DEC} = 0.4\arcsec\pm 0.2\arcsec$.
%For the central region, the offset is negligible.
%This offset was taken into account for any further cross-matching with other catalogues.
To minimise the influence of such an astrometric offset, we additionally check any subsequent cross-matching with other catalogues individually by eye.

\subsection{Spectral Index}
\begin{figure}
    \centering
    \includegraphics[width = \linewidth]{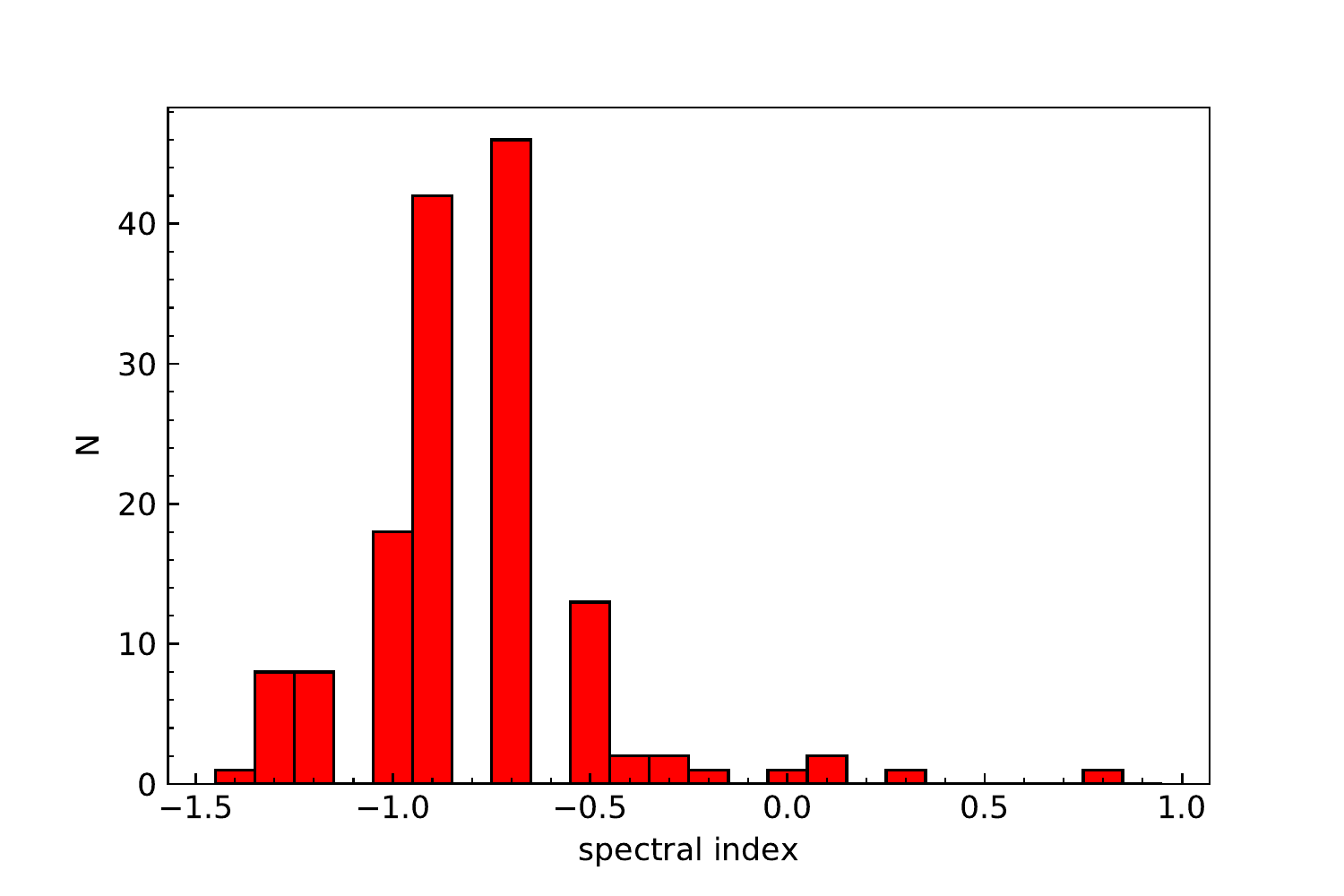}
    \caption{Histogram of the total power spectral indices for our polarised sources. The spectral indices between 150\,MHz and 1.4\,GHz are taken from \citet{Mahony2016}}
    \label{fig:spec_hist}
\end{figure}

To explore the spectral properties of the polarised sources, we used the catalogue of \cite{Mahony2016} who determined the spectral index of sources in the Lockman Hole field using LOFAR 150\,MHz and WSRT 1.4\,GHz observations. 
For crossmatching our sources with their catalogue, we used the coordinates of the sources in total power and a search radius of 15\arcsec. 
The best match for every source was then additionally inspected by eye. 
For four polarised sources no LOFAR counterpart was found.
In Fig. \ref{fig:spec_hist} we show a histogram of the total power spectral indices of our polarised sources. The histogram shows the same source distribution as \cite{Mahony2016} show in their paper (Fig. 15) for their whole Lockman Hole sample. 
The median spectral index of our sample is $\alpha = -0.8 \pm 0.07$ which is consistent, within uncertainties, with the median spectral index from \citet{Mahony2016} of $\alpha = -0.78 \pm  0.015$.
Throughout this publication we define the spectral index $\alpha$ using the relation $S\propto \nu^\alpha$.

\subsection{Infrared cross-identification}
Following the same procedure as presented by \cite{Hales2014b}, we %To distinguish between starforming galaxies (SFGs) and AGN we 
used the far-infrared-radio correlation, plus mid-infrared colours, to distinguish galaxies dominated by star-formation (i.e. star-forming galaxies, SFGs) from AGN. 
%We used the SPITZER Wide-area InfraRed Extragalactic Survey \citep[SWIRE;][]{SWIRE}, which observed huge parts of the
As part of the SWIRE survey the Lockman Hole was observed with the InfraRed Array Camera \citep[IRAC;][]{IRAC} as well as with the Multiband Imaging Photometer for Spitzer \citep[MIPS;][]{MIPS}.
IRAC observes in four mid-infrared bands, with effective wavelengths of 3.6, 4.5, 5.8 and 8.0\,$\mu$m. MIPS observes in the far-infrared (FIR) bands at 24, 70 and 160\,$\mu$m. 
The data were taken from the online NASA/IPAC science archive\footnote{\url{irsa.ipac.caltech.edu/data/SPITZER/SWIRE/}}.
For cross-identification, we first determined the best matching source within a radius of $15\,\arcsec$ of our total intensity sources. 
The cross matches of all polarised sources were than checked by eye for verification, using the polarised and total intensity maps and the SWIRE images.
After this cross-matching and visual inspection, we found $35$ sources were detected in all four IRAC bands, and $14$ of these were also detected at $24\,\upmu$m in MIPS. Four sources were detected at $24\,\upmu$m but not in all IRAC bands.
In total we found SWIRE counterparts for 39 of our polarised sources.

\subsection{Photometric redshift}
In order to investigate the redshift dependence of characteristics within our polarised source population, we cross-matched our sample with the catalogue from \cite{Hendrik}, who used deep five-band optical imaging of the Lockman Hole from CFHT to derive photometric redshifts. 
A direct cross-match between our radio catalogue and the aforementioned optical catalogue from \cite{Hendrik} did not give reliable results, as within one beam we found multiple possible optical counterparts.
To remedy this problem, we sought to use the higher resolution infrared information as a bridge between our optical and radio catalogues. The brightness correlation between optical and radio wavelengths is weak for galaxies, however there is a significant correlation between sources radio and their infrared fluxes. As a result, sources dominating the flux within our radio beam should therefore also be the brightest in the infrared images. Furthermore, the mid-infrared images are sufficiently high-resolution such that optical-FIR cross-matches are unique; both the optical and FIR imaging have resolutions better than 2\arcsec in full-width at half-maximum (FWHM) of the point-spread function (PSF). Therefore, assuming the radio-MIR correlation holds, we are able to uniquely map optical counterparts to our radio emission via the high-resolution SWIRE imaging. 
% of 1.6\arcsec to 1.9\arcsec (full-width at half-maximum of the point-spread function), depending on the band. This is superior to the radio and similar to the optical images, so that we were able to cross-identify the optical sources using this intermediate step. 
Additionally, we visually inspected each source, using our polarised and total intensity map, the SWIRE infrared images, and the optical images, to identify false associations.

Sources which showed clear AGN jet structure in our total intensity map but were not detected by SWIRE were used to co-locate the central host galaxy. This was done by eye and enabled us to cross-identify two additional sources.
%The resolved jet structure allowed us in some cases to identify its origin.
Using this combined strategy we were able to associate optical counterparts from the catalogue of \cite{Hendrik} and thus photometric redshifts to 56 of our polarised sources.

%-----------------------------------------------------------------

\section{Source Classification}
\label{sect_source_classification}

In the following we classify our polarised sources and inspect whether star-formation dominated systems are found in our polarised source sample. For this purpose we used the FIR-Radio-correlation (FRC) and infrared colour-colour diagrams, following the procedure outlined in \cite{Hales2014b}. The polarised AGN sample is then classified into Fanaroff-Riley-types \citep{Fanaroff1974} using morphological parameters and radio and optical brightness values.

\subsection{FIR-Radio-correlation}

\begin{figure}
    \centering
    \includegraphics[width = \linewidth]{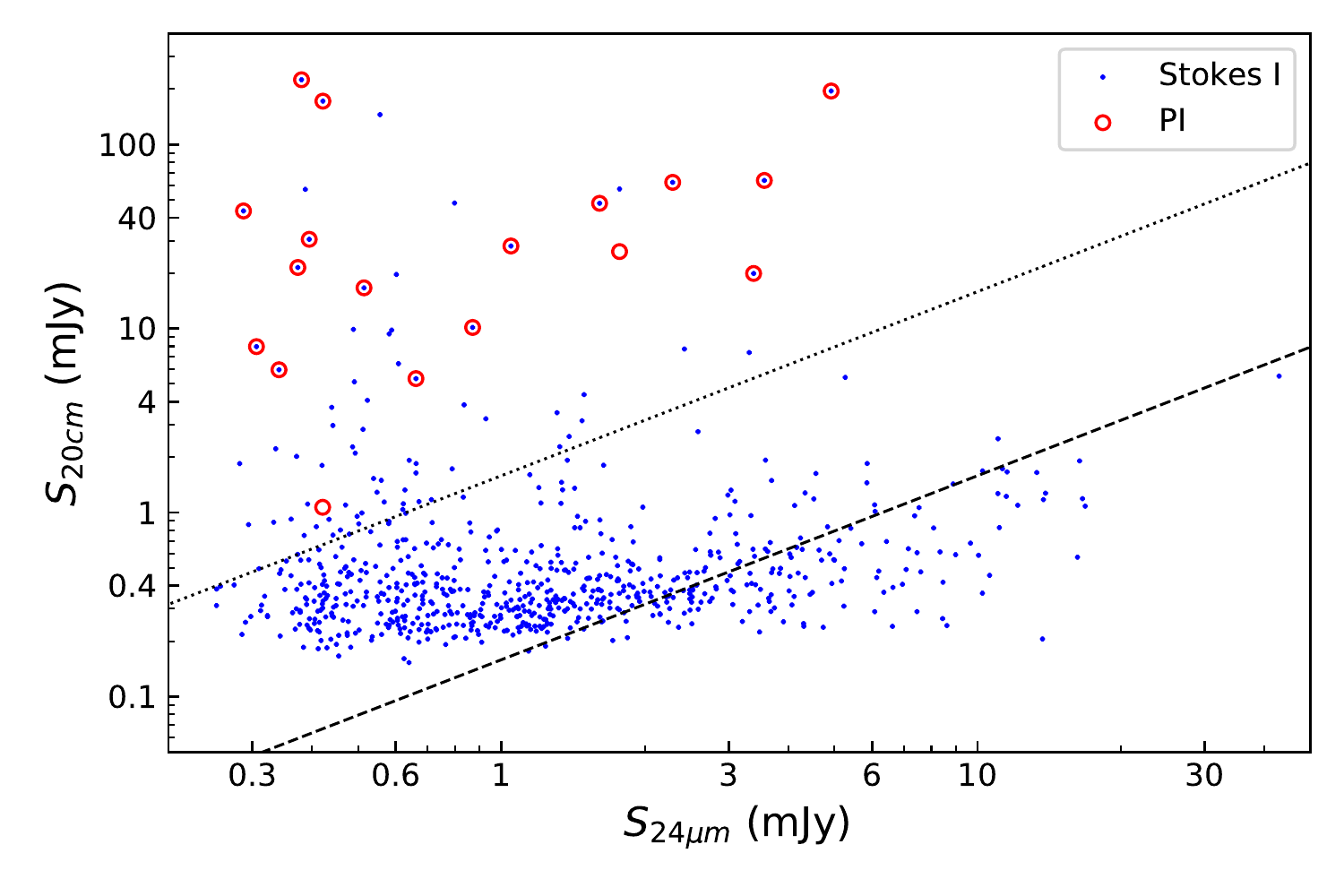}
    \caption{Radio flux density compared to far infrared flux density for all sources with a 24$\,\mu$m counterpart. The red circles indicate the 18 polarised sources with a 24$\,\mu$m counterpart. The dashed line gives the far infrared radio correlation from \cite{Appleton2004}, the dotted line indicates the radio flux density being 10 times greater than for the FRC.}
    \label{FRC}
\end{figure}

The FRC gives a correlation between the far-infrared flux and the radio flux, observed for starforming systems. This correlation is commonly described by the parameter $q_{24}$ which is defined by $q_{24} = \text{log}_{10}[S_{24\,\mu m}/S_{20\,cm}].$
Fig. \ref{FRC} compares the radio flux densities of our 18 polarised sources to their FIR 24$\,\mu$m flux densities from SWIRE. The dashed line indicates the FRC, defined by \cite{Appleton2004} as $q_{24}=0.8$. Following \cite{Hales2014b} we used the dotted line as a criterion for classifying sources as AGN. For this limit the radio flux density is at least 10 times higher than the FRC, so that for these sources $q_{24} \geq -0.2$ holds.
All polarised sources are clearly situated above our set limits and thus are classified as AGN. 

\subsection{Infrared colour-colour diagram}

\begin{figure}
    \centering
    \includegraphics[width = \linewidth]{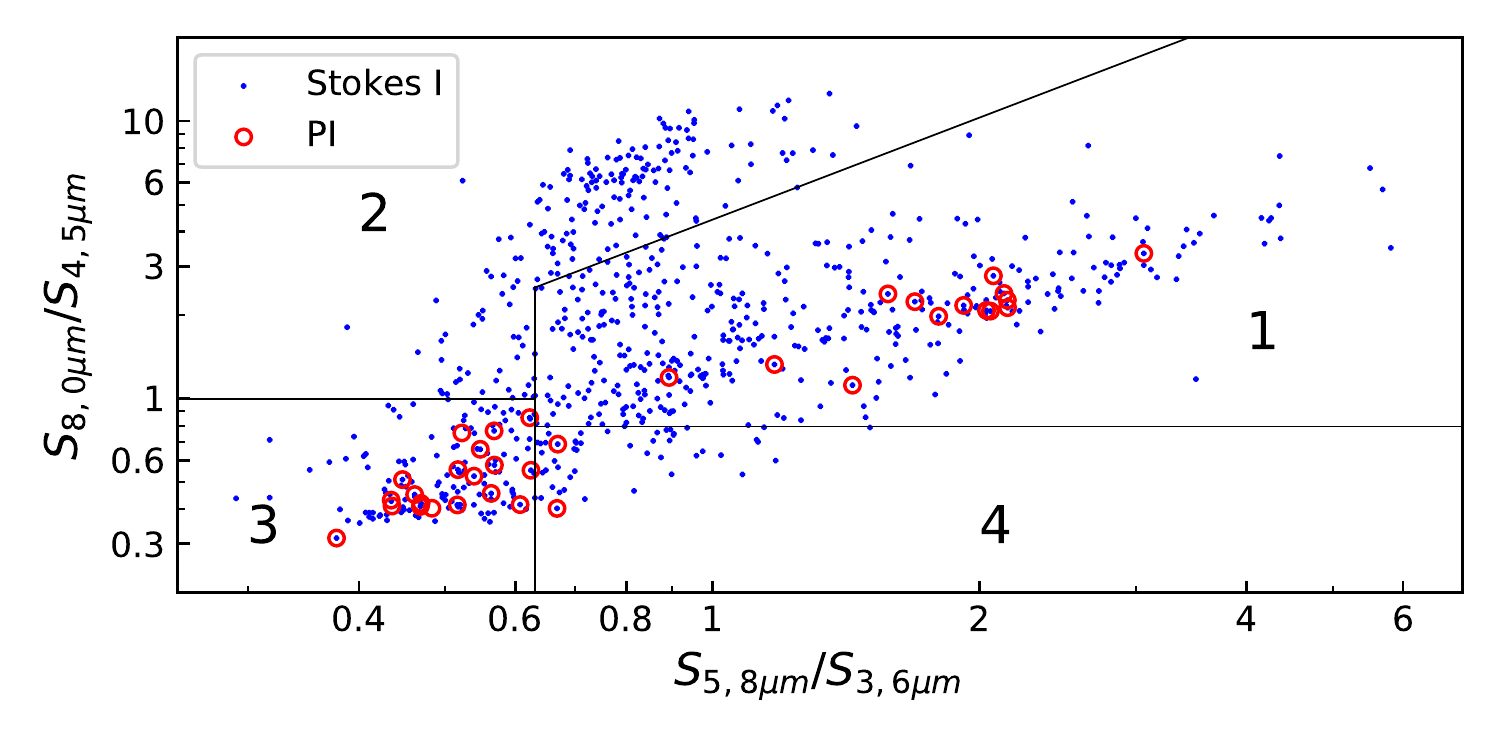}
    \caption{SPITZER mid-infrared colour-colour diagram of the sources in the Lockman Hole Field. The blue dots show all sources for which infrared counterparts could be found. The open red circles indicate the 34 polarised sources detected in all four IRAC bands. For the explanation of the numbered regions see Sect. \ref{sect_source_classification}.}
    \label{IR_color_color}
\end{figure}

Fig. \ref{IR_color_color} shows the SPITZER mid-infrared colour-colour diagram, where the flux density ratios $S_{8.0\,\mu m}/S_{4.5\,\mu m}$ and $S_{5.8\,\mu m}/S_{3.6\,\mu m}$ are compared with each other.
The blue dots represent all our Stokes I sources for which infrared counterparts could be found in all four mid-infrared bands. The red circles represent the polarised sources. 
Following \cite{Sajina2005}, who simulated the diagram for a redshift range of $0 < z < 2$, region 1 hosts mostly continuum dominated sources at $0 < z < 2$ (which is an indication for AGN activity), region 2 selects preferentially poly-aromatic hydrocarbon (PAH) dominated sources at $0.05 < z < 0.3$ (which indicates star-forming galaxies), region 3 and 4 host mainly stellar- and PAH-dominated objects at $0.3 < z < 1.6$ and $z > 1.6$ respectively. For higher redshifts sources migrate from region 2 to region 4. 
Therefore, as sources in region 1 can most likely be identified as AGN, all polarised sources in region 1 are classified as AGN.
Conversely, the radio emission of the sources in region 3 and 4 might originate either from AGN activity, while the infrared colours are dominated by an old stellar population, or from star-formation.
However, since we found no polarised source in region 2 (suggesting no certain star-formation contamination), we conclude that there is no evidence of detection of SFGs in polarised emission within our sample.  
If those were migrated PAH dominated sources from region 2, we would expect to observe at least a few remaining sources at low redshift in region 2 \citep{Hales2014b}. We therefore classified all polarised sources in region 3 and 4 as AGN. No hint for any detection of SFGs in polarised emission was found in the part of our sample with sufficient multi-wavelength data. We assume this part to be representative for our whole sample.%our sample.
Although this was expected, we note that our classification procedure is just a statistical procedure and does not account for variations within individual sources.

\subsection{AGN radio power}
\label{sec:FR-Classification}
 
%\textit{For a more detailed analysis of our polarised AGN sample we used the morphological Fanaroff-Riley (FR) classification scheme \citep{Fanaroff1974} to differentiate between different types of AGN. FR\,I galaxies are core dominated AGN while FR\,II galaxies are lobe dominated. 
%Since most of our objects are unresolved, a classification based on the morphology is only possible for a small fraction of sources.}

To differentiate between weaker and stronger AGN we use a criterion from \cite{Gendre2008} who used the absolute radio luminosity at 1.4\,GHz P$_{1400}$ and the optical B magnitude M$_B$:
%Due to this we used the absolute radio brightness of the sources for classification. 
%For this purpose we classify our sources into FR\,I and FR\,II types by absolute radio luminosity and B-magnitude. 
%\textit{Due to this we used the absolute radio brightness at $1.4\,$GHz $P_{1400}$ and the optical B magnitude $M_B$ from the catalogue of \cite{Hendrik} to differentiate between the FR types. 
%\cite{Gendre2008} showed that a cutoff of}
\begin{equation}
    \text{logP}_{1400} = -0.27 M_B + 18.8
    \label{P-MagB-FR}
\end{equation}
%\textit{produced results where most of the sources below this criterion were assigned to the correct FR\,I phenotype, and sources above this criteria were correctly assigned to the FR\,II phenotype.}
We use the optical B magnitude from the catalogue from \cite{Hendrik}.
This cutoff was formerly used as a criterion for a statistical and non-morphological way to classify AGNs into Fanaroff-Riley (FR) types \citep{Fanaroff1974}. 
\cite{Mingo2019} showed that this criterion does not nessecerily hold for the morphological Fanaroff-Riley types. We thus do not claim our sources to be of the phenotypes FRI and FRII, but differentiate between weaker and stronger radio AGN.
%Even though it was found by \cite{Mingo2019} that the radio luminosity in relation to the host galaxies properties do not give a valid criterion for classifying AGN following the FR-classification, we use it to differentiate between weaker and stronger AGN.}

\begin{figure}
    \centering
    \includegraphics[width = \linewidth]{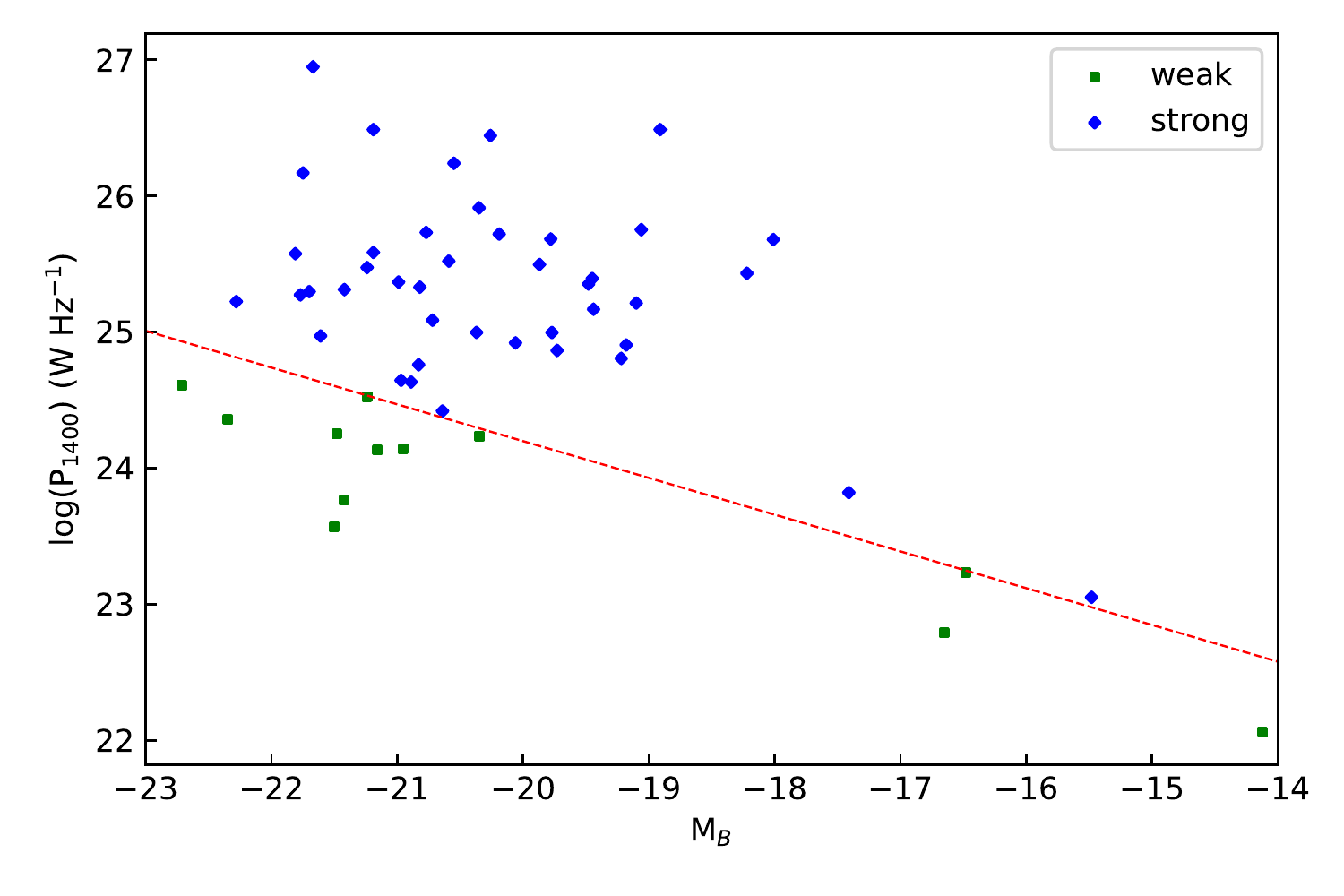}
    \caption{Absolute radio brightness calculated using the photometric redshifts over the B-magnitude for the polarised AGN sources. Green symbols represent sources classified as weak AGN and blue ones sources classified as strong AGN. The dashed line is given by Eq. \ref{P-MagB-FR} and resembles the cutoff previously used to differentiate between the FR phenotypes.}
    \label{fig:FR-class-magB}
\end{figure}

%Although there is not a strong cut between the brighter FR\,II and the fainter FR\,I types, we classified all sources with $\text{log}(P_{1400})>25 \text{ W Hz}^{-1}$ as FR\,II and the fainter ones as FR \,I.
%\textit{To calculate the absolute luminosity we used the equation given by \cite{Ledlow1995}, which accounts for the geometry of the Universe where $H_0=75\text{km s}^{-1}\text{Mpc}^{-1}$ is the Hubble constant and $q_0 = 0$ the present day deceleration parameter:
%\begin{equation}
%    \text{logP}_{1400} = \text{log}(1.91167 \cdot 10^{27}[\text{z}(1+\frac{\text{z}}{2})]^2 \text{I}_{1400}(Jy)) \text{ [W Hz}^{-1}]
%\end{equation}
%$\text{I}_{1400}$ is the observed total intensity of the source.
%Using the redshift information we have for 56 of our polarised sources, we were able to classify 14 FR\,I and 42 FR\,II sources (Fig. \ref{fig:FR-class-magB}).}
We calculated the absolute radio luminosity in the following way:
\begin{equation}
    \text{P}_{1400} = 4 \pi \text{DL}^2 \text{I} (1+z)^{-(\alpha+1)}
\end{equation}
where DL is the luminosity distance calculated using the \textsc{luminosity\_distance} function of the python package astropy.cosmology.FlatLambdaCDM\footnote{\url{https://docs.astropy.org/en/stable/cosmology/index.html}}. We use a Hubble constant $H_0 = 67 \frac{km/s}{Mpc}$, $\Omega_m = 0.31$ and a CMB-Temperature of 2.7K.
I is the observed total intensity of our sources, z is the redshift and $\alpha$ the spectal index. Using the redshift information we have for 56 of our sources we classify 14 weaker and 42 stronger AGN (see Fig. \ref{fig:FR-class-magB}).
The uncertainties shown in Fig. \ref{fig:FR-class-magB} originate from the photometric redshift uncertainties derived from \cite{Hendrik}, who provide minimum and maximum photometric redshift confidence limits in addition to the maximum posterior point estimate. 

%Since all our sources are classified as AGN, we used the redshift information we have for 56 of our sources, to classify them following the Fanaroff-Riley classification. 

%-----------------------------------------------------------------

\section{Results}

\subsection{Selection effects}
\label{sec:selection_effect}

An anti-correlation between the fractional polarisation and the total intensity flux density was observed by \citet{Tucci2004,Mesa2002} and \citet{Grant2010}. This anti-correlation was claimed to be caused by either an increasing median redshift of faint sources \citep{Tucci2004}, or a change in the composition of the observed population \citep{Mesa2002}. \cite{Hales2014b} explained this as a selection effect due to incompleteness of the sample, arguing that, for the faintest sources (in total intensity) that reside close to the detection limit of the observations, it is impossible to detect low fractional polarisations. 

\begin{figure}
    \centering
    \includegraphics[width = \linewidth]{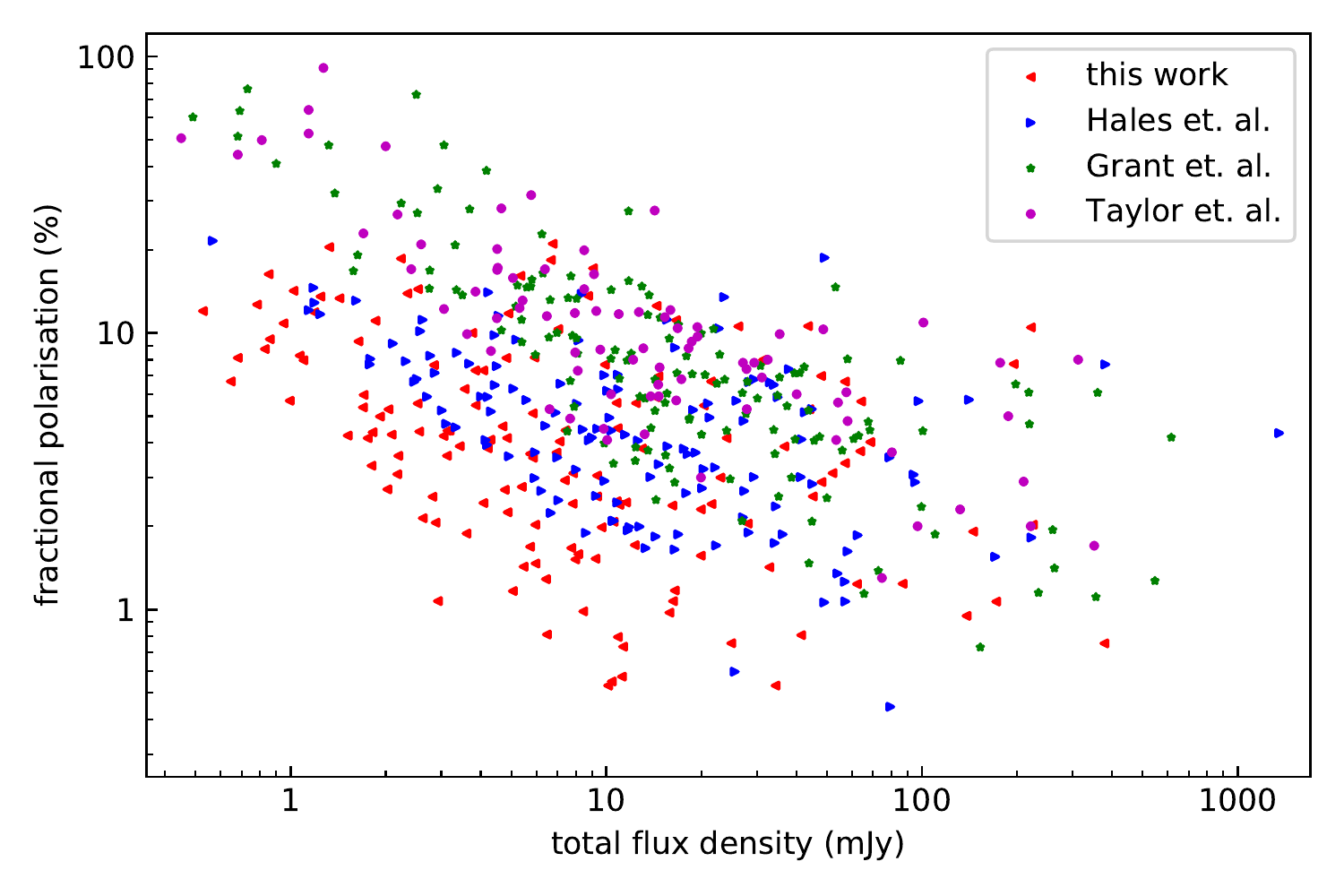}
    \caption{Fractional polarisation over total intensity for sources of different publications. Red triangles looking to the left represent the sources in this work, blue triangles looking to the right the ones from \citet{Hales2014b}, green stars the ones from \citet{Grant2010} and magenta circles the ones from \citet{Taylor2007}.}
    \label{Vergleich_Kataloge_log}
\end{figure}

Fig. \ref{Vergleich_Kataloge_log} shows the fractional polarisation as function of the total intensity for our sources (red), and literature sources from \citet[blue]{Hales2014b}, \citet[green]{Grant2010}, and \citet[magenta]{Taylor2007}. Our sample exceeds the sensitivity of the observations by \citet{Tucci2004,Mesa2002,Taylor2007} and \cite{Grant2010} by roughly a factor of six, and is about twice as sensitive as \citet{Hales2014b}. We are therefore able to reach lower fractional polarisation levels for sources showing total power emission of the same intensity. The theoretical cutoff under which no sources can be found anymore can simply be described by the sensitivity limit PI$_{min}$ of every sample:
\begin{equation}
    \Pi_{min} = \frac{\text{PI}_{min}}{\text{I}}
\end{equation}
For real observations, other effects, such as the reduced sensitivity closer to the borders of the mosaic, reduce this limit and the above equation is only an approximation. 

\begin{figure}
    \centering
    \includegraphics[width = \linewidth]{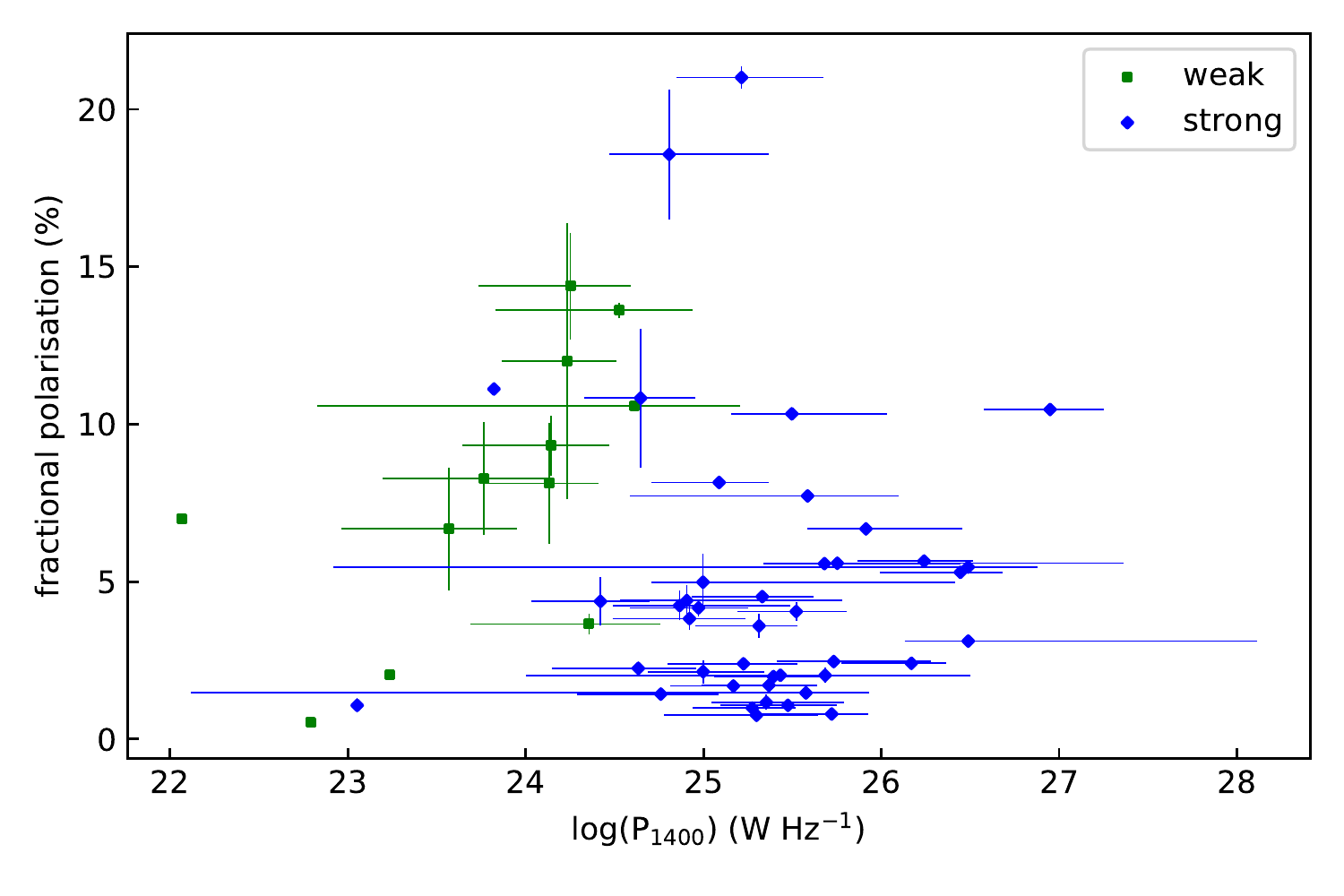}
    \caption{Fractional polarisation over absolute radio brightness for the 56 sources with known photometric redshift.}
    \label{fig:frac_P}
\end{figure}

Since we were able to calculate the absolute radio luminosity of 56 sources we also plotted the fractional polarisation over the absolute radio luminosity in Fig. \ref{fig:frac_P}. 
A physical reason for a correlation between the total flux density and the fractional polarisation should also show up as a correlation between the absolute brightness and the fractional polarisation. Fig. \ref{fig:frac_P} does not show explicit hints of such a correlation. 
This is confirmed by binning the data by their total radio brightness (see table \ref{A:bins_absolutbrightness}). 

We therefore conclude that the observed anti-correlation between fractional polarisation and total intensity has no physical origin and instead, following \citet{Hales2014b}, is caused by a selection effect due to limited sensitivity.

%\begin{figure}
%    \centering
%    \includegraphics[width = %\linewidth]{Bilder/Vergleich_Kataloge_paper.pdf}
%    \caption{Caption}
%    \label{Vergleich_Kataloge}
%\end{figure}

\subsection{Comparison of different Deep Fields}
\label{sec:cosmic_variance}

\begin{figure}
    \centering
    \includegraphics[width = \linewidth]{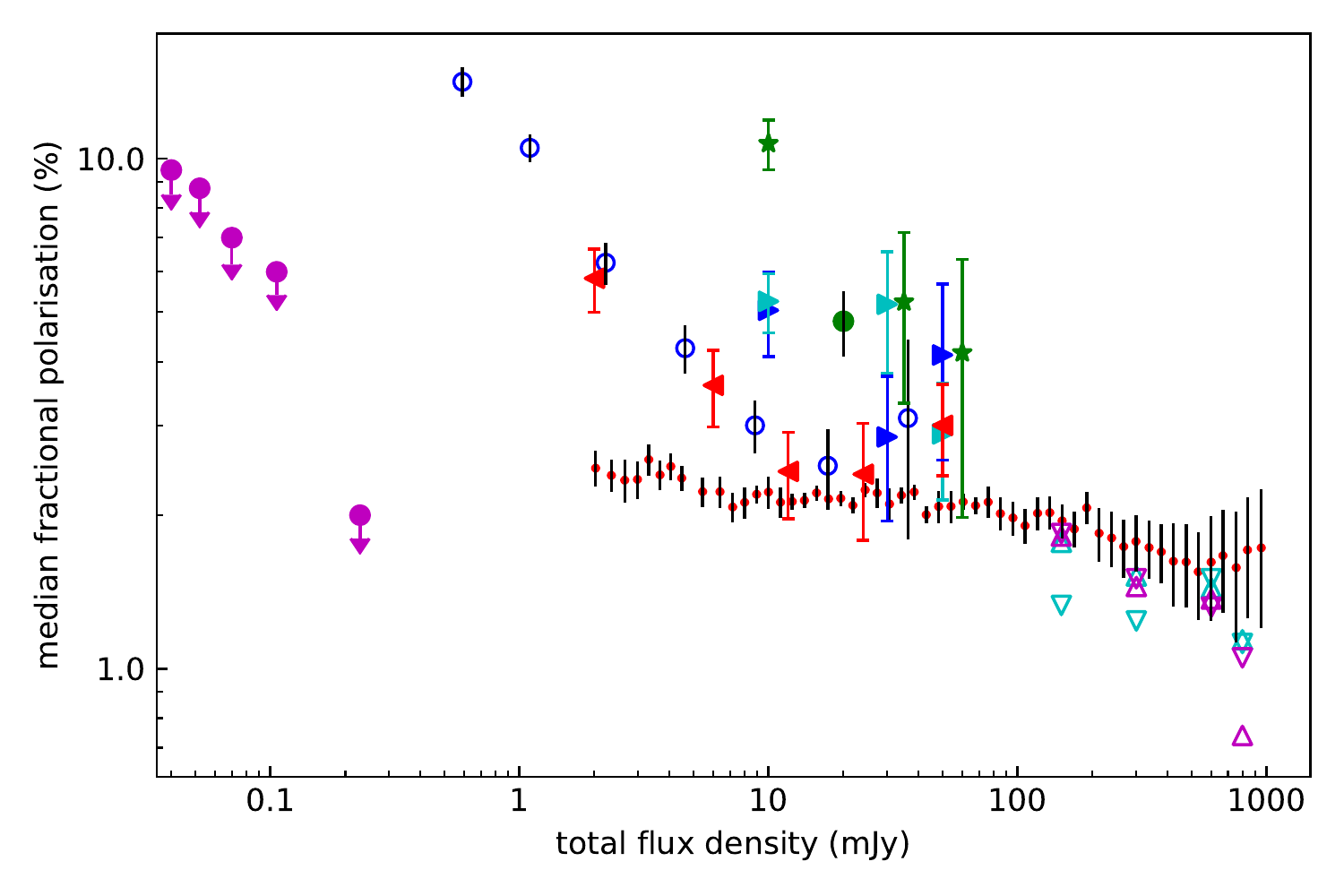}
    \caption{Median fractional polarisation over total intensity, for different catalogues. The red dots give the datapoints of \cite{Stil2014} who used the NVSS catalogue and a stacking technique, the open turquoise up- (steep) and downlooking (flat) triangles are the polarised sources form \cite{Tucci2004} of the NVSS split up by a spectral index of -0.5. The open magenta triangles represent the sources from \cite{Mesa2002} split by the same spectral index limit. 
    The magenta circles represent upper limits from the catalogue of \cite{Rudnick2014}. The green stars are from \cite{Grant2010}, the green circle from \cite{Taylor2007} and the blue open circles from \cite{Subrahmanyan2010}. The triangles facing right are the data from \cite{Hales2014b}, where the blue ones represent their data for the ELAIS-S1 and the turqoise ones their data for the CDF-S Fields.
    The red triangles facing left represent our polarised sample.}
    \label{Stil_plot}
\end{figure}

We now compare the fractional polarisation of our sources with all available literature data to investigate the dependence of the fractional polarisation on the total power emission (Fig. \ref{Stil_plot}). 
For this purpose we binned our data by the total intensity of our polarised sources.  
The bin limits and the respective sizes are listed in Table \ref{table:bins}.
%The first bin contains 60 sources with total intensity fluxes lower than $5\,$mJy. The second bin contains 58 sources where the total flux is in the range of $5 - 20\,$mJy. The third bin contains all 37 sources with fluxes higher than $20\,$mJy.
We have similarly binned the data from \cite{Hales2014b}, for the ELAIS-S1 and the CDF-S fields, and from \cite{Grant2010}, and included them in Fig. \ref{Stil_plot} for comparison. We also list the counts per bin for these sources in Table \ref{table:bins_others}.
%we chose are $0.49-20\,$mJy, containing 28, 56 and 81 sources, $20-40\,$mJy containing 10, 14 and 31 sources and sources with total intensities higher $40\,$mJy containing 7, 15 and 24 sources, respectively.
Additional literature data shown in Fig. \ref{Stil_plot} are taken directly from the publications.
The data from \cite{Stil2014} are generated using a stacking technique. %, this should be a good approximation for a complete sample. 
%For a bin with high completeness we would thus expect the fractional polarisation to fit those data. It is visible that is is barley true for any sample, with exception for the data from \cite{Mesa2002} and \cite{Tucci2004}, who also used the NVSS.
With the exception of \citet{Stil2014}, \citet{Mesa2002}, and \citet{Tucci2004}, all publications used in Fig. \ref{Stil_plot} utilise deep observations of small fields and thus only sample a small part of the sky. 
An overview of these fields areas and sensitivities is given in Table \ref{A:overview}. 
For the bright sky, down to fluxes of $\sim$ 60\,mJy, only NVSS-based data is available in an amount that is  usable for statistical comparison. For the fainter sky there are no direct NVSS observations, but instead only the stacked results from \citet{Stil2014}.
Comparing the small area samples with each other reveals different trends in the median fractional polarisation of the faint polarised radio sky. %50\,mJy.
The sample used here, along with that from \cite{Subrahmanyan2010} and the ELAIS-S1 sample from \cite{Hales2014b}, shows an increase in median fractional polarisation to both lower and higher total intensities, around a minimum. On the other hand the median fractional polarisation values of the sample of \cite{Rudnick2014} and the CDF-S observations by \cite{Hales2014b} increase and finally flatten towards low flux densities. It is important to note, though, that the analysis by \cite{Rudnick2014} only presents upper limits, which are mainly driven by completeness limitations.
However, for the same fluxes the median fractional polarisation of the different fields differs by a factor of 2 or more.
In general an increase of median fractional polarisation towards low flux densities may be due to the aforementioned completeness limitations. This is, however, in contrast to the observed flattening towards low flux densities. In addition, since all analysis of the faint polarised sky, with the exception of \citet{Stil2014}, are based on areas of only several square degrees, sample variance of the underlying large scale structures (i.e. variation driven by the existence of superclusters and voids in the fields) are affecting the results. We can therefore conclude that the characteristics of the polarisation properties of the faint sky below 60\,mJy are not properly understood and that deeper, wider observations are needed to increase the completeness of these samples.

\begin{table}[]
\centering
\caption{Overview of the bins used for Fig. \ref{Stil_plot} for the data of \cite{Hales2014a} and \citet{Grant2010}.}
\label{table:bins_others}
\begin{tabular}{c|c|c|c}
bin & N$_{ELAIS-S1}$ & N$_{CDF-S}$ & N$_{ELAIS-N1}$ \\
mJy & & &\\
\hline
0.49-20 & 28 & 56 & 81\\
20-40 & 10 & 14 & 31\\
>40 & 7 & 15 & 24\\
\end{tabular}
\tablefoot{The first column gives the flux range, the second the total number of sources in the bin (N) for the ELAIS-S1 field, the third for the CDF-S field both from the observation of \cite{Hales2014a} and the fourth for the ELAIS-N1 field from the observation of \citet{Grant2010}.}
\end{table}

\subsection{Spectral Index}
\label{sec:Spectral_results}

%Following \cite{Hales2014b} we can explain the anti-correlation between fractional polarisation and total intensity flux density as a completeness effect. However \cite{Tucci2004} detected this anti-correlation mainly for steep spectrum sources ($\alpha < -0.5$) but only barely for flat (and inverted) spectrum ($\alpha\geq-0.5$) sources.

\citet{Tucci2004} detected an anti-correlation between the fractional polarisation and the total intensity flux (like already discussed in Sec. \ref{sec:selection_effect}), predominantly for steep spectrum sources ($\alpha < -0.5$) but only marginally for flat (and inverted) spectrum ($\alpha\geq-0.5$) sources.
This observation can not be explained by the selection effect argument of \citet{Hales2014b}. Such an effect would affect steep as well as flat spectrum sources.
In contrast to the above findings, \citet{Mesa2002} did not find such a strong dependency on the spectral index as \citet{Tucci2004}.

\cite{Stil2015} used stacking to increase the sensitivity and thus fill the gap between our data and the NVSS data. They observe only a weak increasing trend for steep ($\alpha < -0.75$) and no increase for flat ($\alpha > -0.3$) spectrum sources, but a strong increasing trend in fractional polarisation for intermediate ($-0.75 < \alpha < -0.3$) spectrum sources.

\begin{figure}
    \centering
    \includegraphics[width = \linewidth]{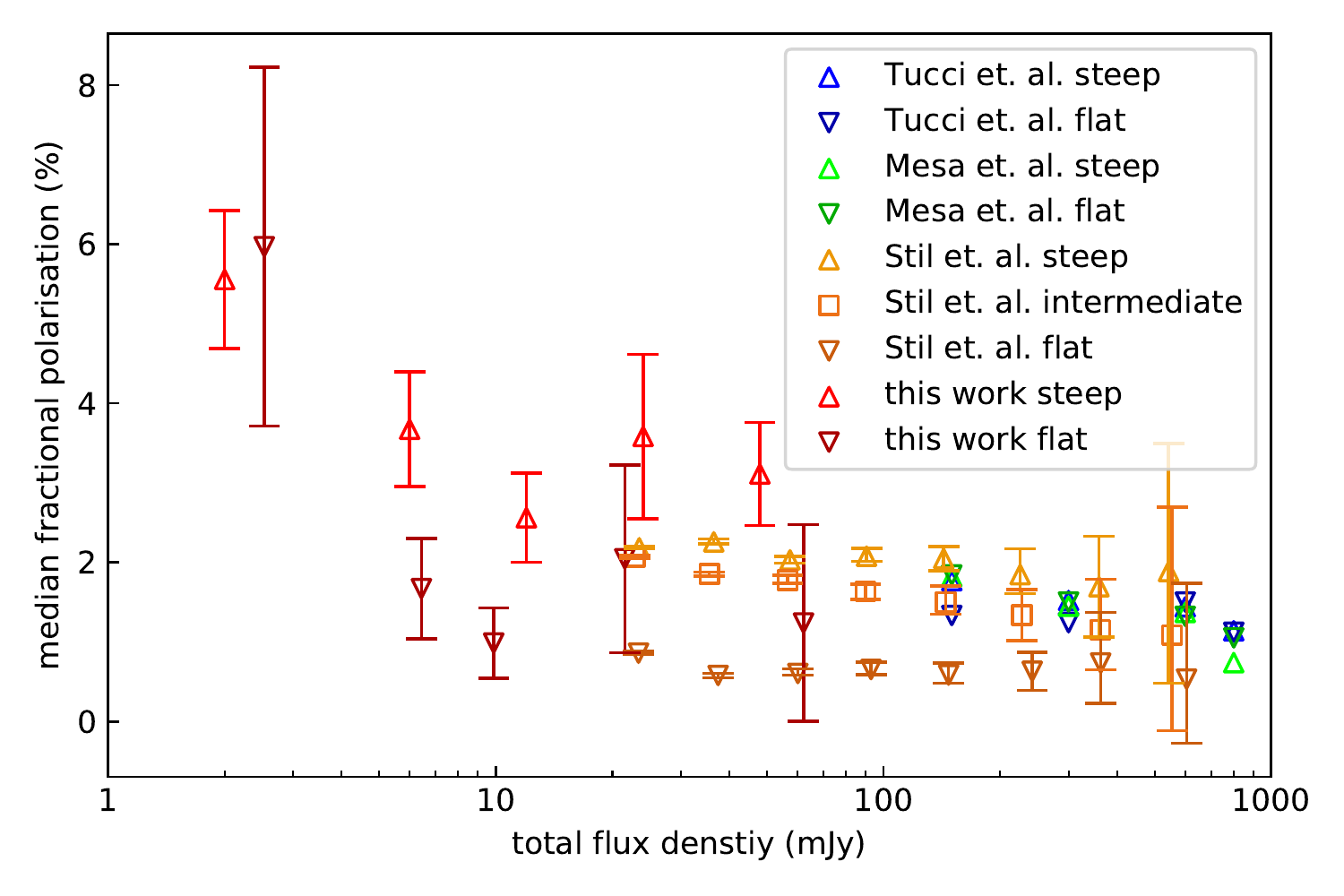}
    \caption{Median fractional polarisation over total intensity for steep and flat spectrum sources for our sample (red), and the ones from \cite{Tucci2004} (blue) and \cite{Mesa2002} (green). Upwards pointing triangles represent steep spectrum sources with a spectral index of $\alpha\leq -0.5$, downwards facing triangles represent flat or inverted spectrum sources with a spectral index of $\alpha\geq -0.5$.}
    \label{Spec_plot}
\end{figure}

In Fig. \ref{Spec_plot} we show the median fractional polarisation for steep and flat spectrum sources respectively, as a function of total flux density.
The number of sources in each flux bin are given in Table \ref{table:bins}.

\begin{table}[]
\centering
\caption{Overview of the bins used for Fig. \ref{Stil_plot} and fig. \ref{Spec_plot}.}
\label{table:bins}
\begin{tabular}{c|c|c|c|c|c|c}
bin & N & $\Pi_m$ & N$_s$ & $\Pi_{m,s}$ & N$_f$ & $\Pi_{m,f}$ \\
mJy & & $\%$ & & $\%$ & & $\%$\\
\hline
< 4 & 50 & 5.83\small{$\pm$0.83} & 41 & 5.56\small{$\pm$0.87} & 7 & 5.97\small{$\pm$2.26} \\
4 - 8 & 34 & 3.59\small{$\pm$0.61} & 26 & 3.67\small{$\pm$0.72} & 7 & 1.67\small{$\pm$0.63} \\
8 - 16 & 27 &  2.44\small{$\pm$0.47} & 21 & 2.56\small{$\pm$0.56} & 5 & 0.98\small{$\pm$0.44} \\
16 - 32 & 15 & 2.41\small{$\pm$0.62} & 12 & 3.58\small{$\pm$1.03} & 3 & 2.04\small{$\pm$1.18} \\
> 32 & 24 & 3.00\small{$\pm$0.61} & 23 & 3.11\small{$\pm$0.65} & 1 & 1.24\small{$\pm$1.24}\\ 
\end{tabular}
\tablefoot{The first column gives the flux range, the second the total number of sources in the bin (N) and the third column gives the median fractional polarisation of the bin ($\Pi_m$) with poisson error. Column four gives the number of steep spectrum sources (N$_s$), column five the median fractional polarisation of the steep spectrum sources per bin ($\Pi_{m,s}$), the sixth and seventh columns give those values for the flat spectrum sources (N$_f$ and $\Pi_{m,f}$).}
\end{table}

%The last bin of the flat spectrum sources needs to bee excluded from the analysis due to the low number of sources.
The last two bins of the flat spectrum sources need to be excluded from the analysis since they contain a statistically insignificant number of sources: 3 and 1, respectively.
It is visible that steep spectrum sources in general have a higher fractional polarisation than flat spectrum sources.
Both types scatter in the same way around a constant value of about 3.25\,\% for steep and 1.62\,\% for flat spectrum sources. They show no sign of correlation (nor anti-correlation) between fractional polarisation and total intensity flux density. 
Only for the bin containing sources with a total flux lower than 4\,mJy the fractional polarisation increases rapidly. 
Since the completeness effect has the strongest influence near the detection limit, this is most likely not caused by physical reasons but rather due to the low completeness in this bin.

%We also checked for this by recalculating the spectral index for all sources using our measured total flux density and the 150\,MHz flux from \cite{Mahony2016}. This ended up with 

We note that the scatter for our sample is intrinsically higher in comparison to the literature values due to the smaller number statistics, when compared to the large samples used for the NVSS analysis.

\subsection{Euclidean normalised polarised differential source counts}
\label{sec:source_counts}

\begin{figure}
    \centering
    \includegraphics[width = \linewidth]{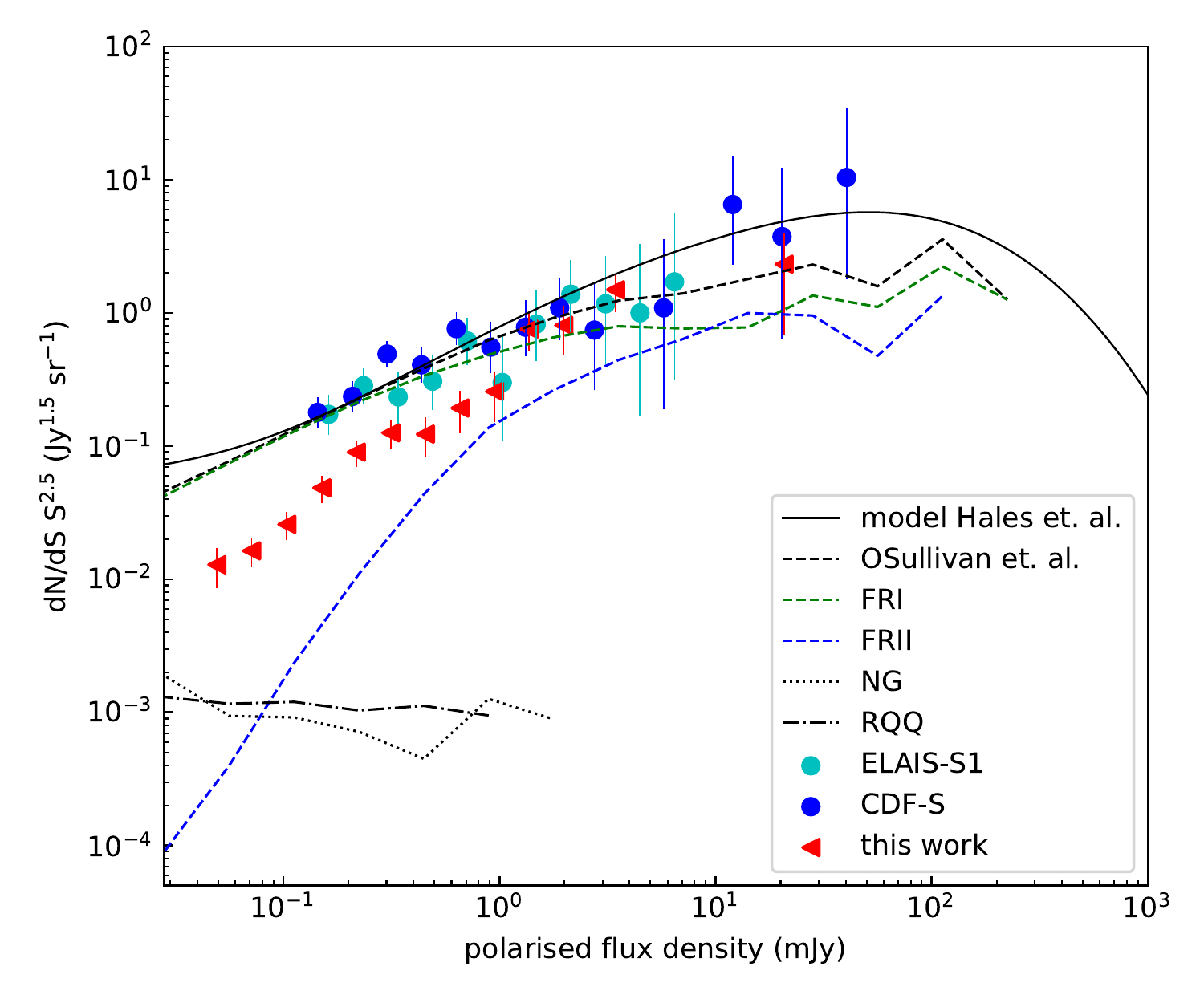}
    \caption{Euclidian-normalized differential source counts at 1.4\,GHz in polarised intensity from our Lockman Hole observations (red triangles). Component counts from \cite{Hales2014b} for the CDF-S (blue) and the ELAIS-S1 (turquoise) are given for comparison. The solid black line shows the model prediction from \cite{Hales2014b}. The other lines are from the simulations of \cite{O'Sullivan2008}; the black dashed line gives the total polarised differential source counts, the green dashed line gives the FR\,I source counts and the blue one the FR\,II source counts. The black dotted line represents the normal galaxies (NG), and the black dot-dashed line the radio quiet quasars (RQQ).}
    \label{fig:diff_source_counts}
\end{figure}

In Fig. \ref{fig:diff_source_counts} we present the euclidean normalised polarised differential source counts from our sources in the Lockman Hole field.
The calculation is based on the code from \cite{Noelia2018}, that accounts for the higher noise at the mosaic edges.
%respects smaller sky areas for lower flux densities due to the higher noise at the mosaic edges. 
We also accounted for the resolution bias, following the method of \cite{LH_Total}.
The solid black line gives the model for polarised differential component counts given by \cite{Hales2014b}, who used the model for total intensity source counts from \cite{Hopkins2003} and convolved it with a polarised density function fitted to their data.
The blue and cyan dots are the data from \cite{Hales2014b} for the CDF-S and the ELAIS-S1 fields. The red triangles represent the polarised differential source counts derived from our sample. For direct comparison, we use the same binning as \cite{Hales2014b} for the CDF-S Field. Following their binwidth of 0.16\,dex, we extended these bins to lower fluxes, excluding bins with sources that are only detectable in a region of less than 10\,\% of our mosaic. We note that the model from \citet{Hales2014b} was calculated for component rather than source counts, but the authors claim that there is no significant difference between source and component counts for resolutions of ~10\arcsec \ in the millijansky regime and below. However it is likely that source counts are slightly lower than component counts, especially when comparing different resolutions.
The dashed black line gives the differential polarised source counts from \cite{O'Sullivan2008}, who used semi-empirical simulations from the European SKA Design Study (SKADS). Using the luminosity they also distinguished between FR\,I and FR\,II sources, for which the source counts are given by the green (FR\,I) and blue (FR\,II) dashed line. The black dotted line represents their source counts for normal galaxies (NG) and the dot-dashed line represents the source counts for radio quiet quasars (RQQ). 

With our study we confirm the decreasing trend for fainter source number counts, while also finding an offset (of up to a factor of 3) towards smaller number counts, over the whole flux range, when comparing our data to those of \citet{Hales2014b}. 
We note, though, that our source counts at low flux densities show a similar trend compared to the FR\,II source counts from \cite{O'Sullivan2008}, even though they are higher. %are similar to the FR\,II source counts from \cite{O'Sullivan2008}. 
Several different reasons can possibly explain this behaviour.
First of all we note that a small offset to the model from \cite{Hales2014b} was expected since we used their fractional polarisation distribution function, which might not properly represent the distribution of our data and component counts might deliver slightly higher values than source counts. %Like we already noted component counts with a resolution of 10'' might be higher than source counts taken from an image with a resolution of 15'', so a offset was expected.
In addition, sample variance introduced by large cosmological structures like superclusters and voids influence the composition of the source counts significantly for surveys spanning only modest (i.e. few square degree) areas on-sky. We expand on the relevance of sample variance in Sec. \ref{sec:discussion}.
%This might be due to the influence caused by cosmic variance, as well as the in sec. \ref{sec:cosmic_variance} presented differences in source characteristics of different small area surveys.
The lower source counts compared to the simulations of \cite{O'Sullivan2008} and \cite{Hales2014b} mean, that we find a dearth of faint sources in our sample. This will be further discussed in Sect. \ref{sec:source_counts_discussion}.

\subsection{Redshift and fractional polarisation}
\label{sec:Redshift_results}

We used the photometric redshift information we have for 56 of our sources to compare the redshift with the fractional polarisation. To this end we binned our sources over five redshift ranges the corresponding numbers are given in table \ref{table:z-bins}.

\begin{table}[]
\centering
\caption{Overview of bins and numbers used for fig. \ref{Redshift} and fig. \ref{Redshift_FR}.}
\label{table:z-bins}
\begin{tabular}{c|c|c|c|c|c|c}
z & N & $\Pi^{med}$ & N$_{weak}$ & $\Pi^{med}_{weak}$ & N$_{strong}$ & $\Pi^{med}_{strong}$ \\
\hline
0.15 & 9 & 6.98\small{$\pm$2.33} & 5 & 6.98\small{$\pm$3.12} & 4 & 4.87\small{$\pm$2.44}\\
0.45 & 9 & 6.68\small{$\pm$2.23} & 6 & 8.80\small{$\pm$3.59} & 3 & 1.42\small{$\pm$0.82}\\
0.75 & 21 & 4.40\small{$\pm$0.96} & 2 & 10.06\small{$\pm$7.11} & 19 & 4.37\small{$\pm$1.00}\\
1.15 & 13 & 3.59\small{$\pm$1.00} & 1 & 0.97\small{$\pm$0.97} & 13 & 3.59\small{$\pm$1.00}\\
1.4 & 4 & 1.94\small{$\pm$0.97} & 0 &  & 3 & 2.41\small{$\pm$1.39}\\
\end{tabular}
\tablefoot{The first column gives the central redshift of the bin, the second gives the total number of sources in the bin (N), the third the median fractional polarisation of the bin ($\Pi^{med}$). The columns four and five give the number of weak sources (N$_{weak}$) and their median fractional polarisation ($\Pi^{med}_{weak}$), while the columns six and seven give the same values for the strong sources.}
\end{table}

\begin{figure}
    \centering
    \includegraphics[width = \linewidth]{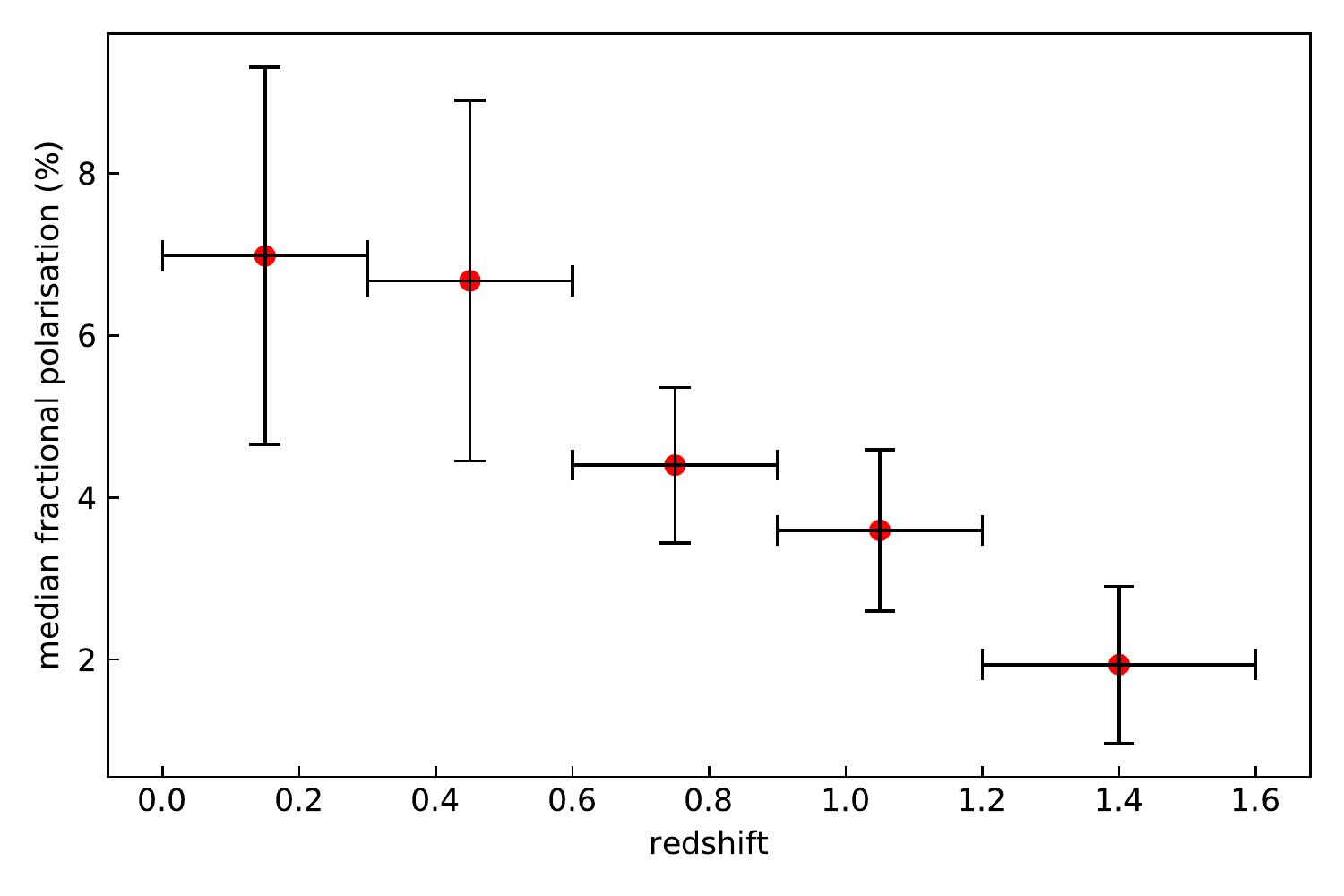}
    \caption{Median fractional polarisation over redshift. The errorbars of the median fractional polarisation are the Poisson errors, whereas the errorbars in x-direction give the bin width.}
    \label{Redshift}
\end{figure}

In Fig. \ref{Redshift} we show the median fractional polarisation as a function of redshift. 
The given uncertainties are calculated assuming Poisson statistics. We find a decrease in median fractional polarisation towards higher redshift ranges. 

An anti-correlating trend was already found by \cite{2012arXiv1209.1438H} for an NVSS polarised sample. They explained it to be due to different types of sources being dominantly observed at different redshifts. Where their low redshift sources are mostly lobes of galaxies with a high median fractional polarisation they found core dominated quasars with a low fractional polarisation at higher redshifts. They found no such trend for either type only. 
In contrast to our study they found a strong decrease of fractional polarisation and than a flattening. 
Since we do not have such a clear turn off, but a decrease over the full sampled redshift range, and no other hint for different source types dominating different redshift ranges, this can not be the reason for our observed anti-correlation.
We note, that in contrast to our study, the study of \cite{2012arXiv1209.1438H} observe the bright polarised sky with a beam about three times larger than our beam. 

We first investigate whether we find selection effects in our data, such as those discussed in Sec. \ref{sec:selection_effect}. For this effect, in order to explain the anti-correlation between fractional polarisation and redshift, a correlation between redshift and observed total flux density would be necessary. Therefore, the observed flux would need to increase with increasing redshift. We find no such correlation, and thus cannot explain the anti-correlation between fractional polarisation and redshift as being due to selection effects.

%FR\,II galaxies are usually more luminous than FR\,I galaxies, and we know that the average integrated fractional polarisation is higher for FR\,I than for FR\,II objects \citep[the latter of which exhibit 4\,\% polarisation generally][]{1988ARA&A..26...93S}. 
%This also holds for our sample, where the median fractional polarisation is $8.2\%$ for FR\,I sources, and $3.94\%$ for FR\,II sources. The two above mentioned effects can easily cause a bias towards one or the other FR type for the high or low redshift bins. Therefore, we want to filter our statistics by differentiating between the two FR phenotypes, using the absolute radio brightness to B-magnitude correlation as described in Sect. \ref{sec:FR-Classification}.
It is known, that the different Fanaroff-Riley phenotype also show different polarisation behaviour. FR\,I objects show in average a higher integrated fractional polarisation compared to FR\,II sources \citep[the latter of which exhibit 4\,\% polarisation generally][]{1988ARA&A..26...93S}.
Also are FR\,II sources known to be usually more luminous than FR\,I sources. These to effects can easily cause a bias towards one or the other FR type, for the high or low redshift bins.
Even though we are not able to investigate this bias, we want to filter our statistics by differentiating between weak and strong sources (see Sec. \ref{sec:FR-Classification}) to investigate the effect of the brightness of the sources on the observed anti-correlation.

\begin{figure}
    \centering
    \includegraphics[width = \linewidth]{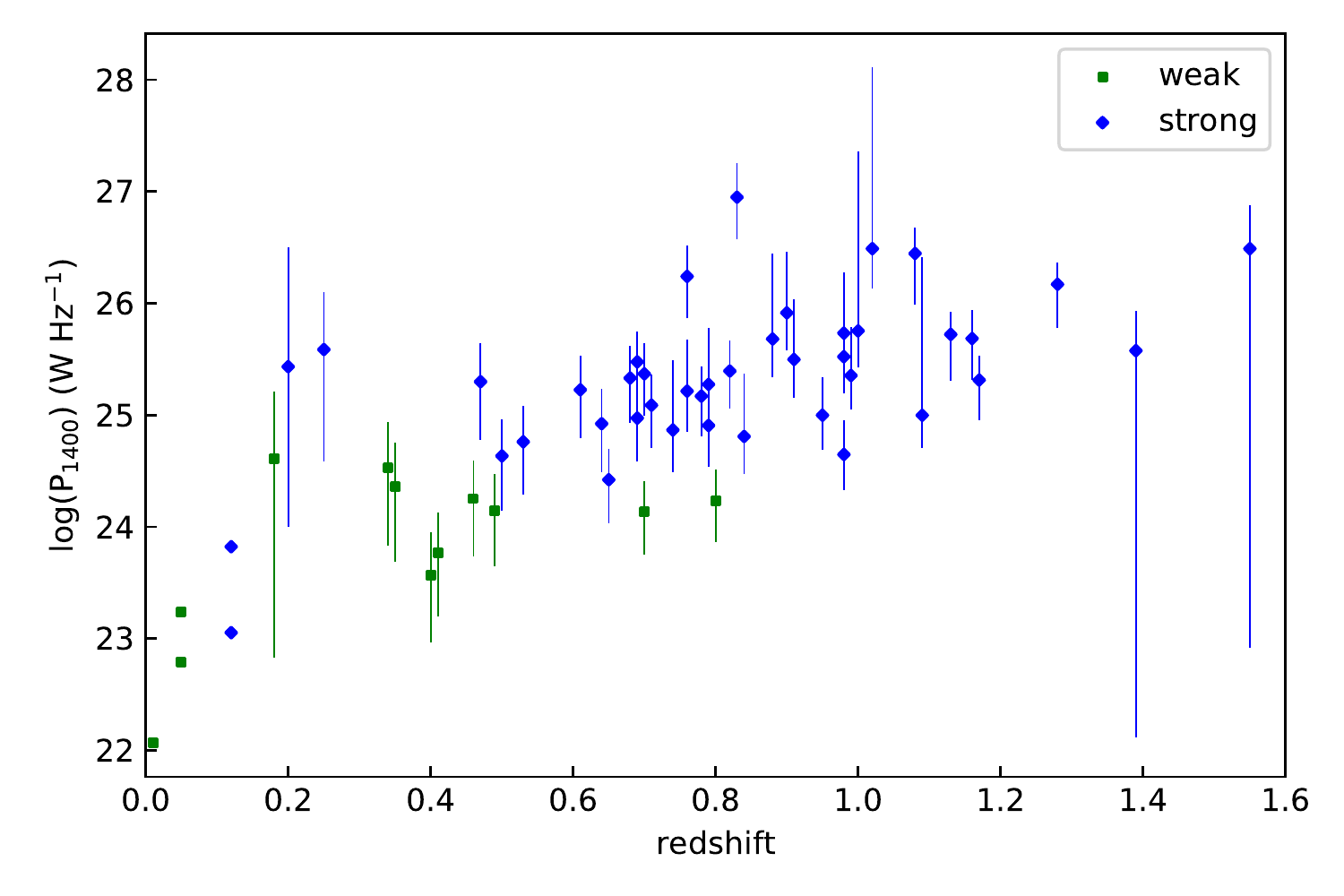}
    \caption{Absolute radio brightness calculated using the photometric redshifts over the redshift. The errors take only the uncertainties of the photometric redshifts into account.}
    \label{FR-class-redshift}
\end{figure}

Fig. \ref{FR-class-redshift} shows the absolute radio brightness of our 56 polarised sources used in this analysis as a function of their photometric redshift. 
With increasing redshift we observed fewer faint and more bright sources. Since the brightest radio sources are (in general) quite rare, it is more likely to observe them with increasing volume \citep[and thus with increasing redshift][]{Ledlow1996}.
The loss of faint sources at high redshift can be explained by selection bias, as fainter sources at high redshift will naturally fall below our detection limit.
%Although FR\,I sources have in general a higher fractional polarisation, they are intrinsically much fainter than FR\,II sources, ending up at lower polarised flux densities. This makes FR\,II sources more likely to be detected in polarised intensity at high redshift than FR\,I sources at comparable redshifts.
%It is also due to this selection effect that we detected more FR\,II sources than FR\,I sources, despite  FR\,I sources being expected to be more common \citep{Ledlow1996}. 

Fig. \ref{FR-class-redshift} shows that our sample is dominated by weak sources for low z and by strong sources for high z.
The last two bins in Fig. \ref{Redshift} are free of any influence from weak sources.
We therefore conclude that the increase of bright sources with redshift cannot alone induce the anti-correlation between fractional polarisation and redshift, given our demonstration (in Fig. \ref{fig:frac_P}) that there is no correlation between fractional polarisation and absolute radio brightness for our sources.

\begin{figure}
    \centering
    \includegraphics[width = \linewidth]{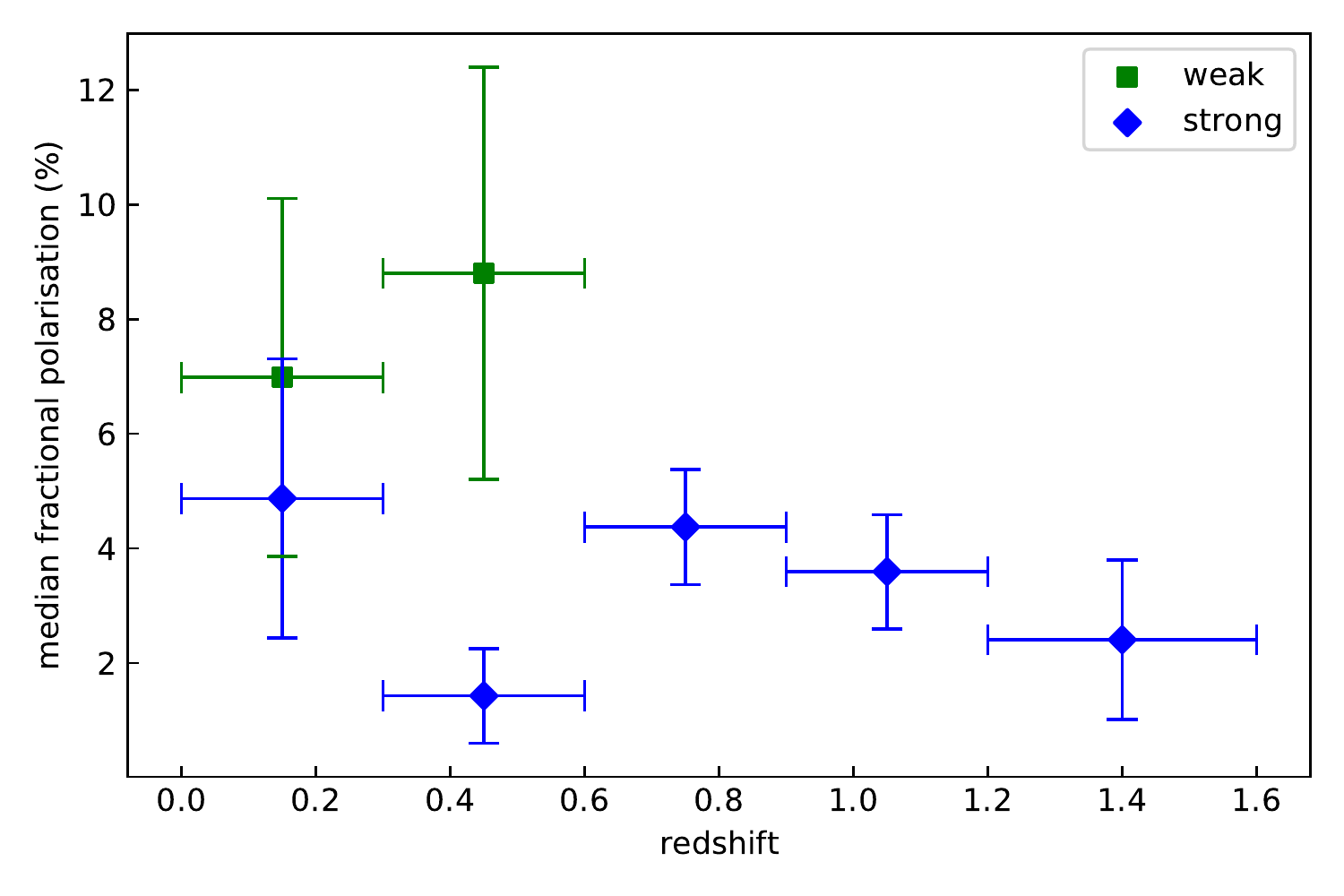}
    \caption{Median fractional polarisation over redshift for weak and strong sources. We excluded those bins only containing less than three sources. In particular that affects the third bin of weak sources.The errorbars of the median fractional polarisation are the Poisson errors, whereas the errorbars in redshift give the binsize}
    \label{Redshift_FR}
\end{figure}

In Fig. \ref{Redshift_FR} we again show the median fractional polarisation over a given redshift range, separated into weak and strong sources. The bins are chosen using the same limits as previously with steps of $\Delta z = 0.3$ for the first four bins and  $\Delta z = 0.4$ for the last bin (see Table \ref{table:z-bins}). 

Since the whole sample contains only 56 sources, it is worth noting that the number of sources in individual bins is small. 
Only the first two bins of weak sources contain a statistically significant number of sources, so we opt to exclude the latter bins from our subsequent analysis. 
%For FR\,II sources the first two bins contain a statistically insignificant number of sources and are thus also excluded from our subsequent analysis (and omitted from Fig. \ref{Redshift_FR}).

%Due to these small number statistics we cannot directly compare the FR\,I and FR\,II polarisation behaviour over redshift, since none of our bins contains a significant number of sources for both types. 
%However, we are able to analyse the behaviour over redshift for the two individual types independently. 
The weak low redshift sources show an increase in median fractional polarisation with increasing redshift. 
We note that only 12 sources contribute to this statistic within two bins, so that the significance is limited. 
%The remaining three bins of the FR\,II sources show a decrease of fractional polarisation with increasing redshift. 
The strong sources show a decreasing trend of fractional polarisation over the whole redshift range. Only the second bin shows a rapid drop. Since only three sources contribute to this bin, we assume it to be an outlier and not of statistical significance. 

%-----------------------------------------------------------------

\section{Influence of cosmic evolution}
\label{sec:redshift-discussion}

The observed anti-correlation between median fractional polarisation and redshift (Sec. \ref{sec:Redshift_results}) can have several origins, which we will discuss individually in this section.

%\subsection{Faraday Dispersion}

First we want to consider the depolarisation of the emission by turbulent magnetic cells, the effect known as Faraday Dispersion.
%the medium in between the source and the observer as the origin. Polarised emission of sources further away cross a larger  volume of magnetoionic medium. 
%This increases the chance for the emission to encounter turbulent magnetic cells in the intergalactic space on the line-of-sight, which cause depolarisation due to Faraday Dispersion.  
%The reason for this anti-correlation might be the longer distance between us and the source.

The ratio $DP$ between the initial polarised flux and the observed polarised flux is strongly wavelength dependent. We also must distinguish between two distinct forms of Faraday Dispersion: the external (EFD) and the internal (IFD) cases \citep{Sokoloff1998}. 
For AGN-sources Faraday dispersion is dominated by external fields situated between the emitting source and the observer.
%In IFD, the depolarising turbulent magnetic field is situated inside the emitting source itself. $DP_{IFD}$ can then be calculated as:
%\begin{equation}
%    DP_{IFD} = \frac{1-\exp{(-2 \sigma^2 %\lambda^4)}}{2\sigma^2 \lambda^4},
%    \label{eq:DP_IFD}
%\end{equation}
%where $\lambda$ is the wavelength.
%In EFD the depolarising turbulent magnetic field is along the line-of-sight between the emitting source and the observer. In this case, $DP_{EFD}$ can be calculated by:
In this case the depolarisation can be calculated by:
\begin{equation}
    DP_{EFD} = \exp{(-2\sigma^2\lambda^4)}.
   \label{eq:DP_EFD}
\end{equation}
%Assuming no difference in source characteristics at different redshifts, we end up with $\sigma$ 
For both cases, $\sigma$ is the same and is defined as:
\begin{equation}
    \sigma ^2 = (0.81\langle n_e \rangle \langle B_{turb} \rangle )^2 Ld/f
    \label{eq:sigma_square}
\end{equation}
%being statistically constant for our case, so that $DP_{IFD}$ is only dependent on the emitted wavelength in the sources restframe.
where $\langle n_e\rangle$ is the average electron density in cm$^{-3}$, $\langle B_{turb} \rangle$ is the average turbulent magnetic field strength in $\mu$G, $d$ is the turbulent cell size in pc, $L$ is the path-length in pc, and $f$ is the filling factor of the turbulent cells causing the depolarisation.

%The ratio between the initial polarised flux and the observed one $DP$ is strongly wavelength dependent. We have to distinguish between two sorts of Faraday Dispersion, the external (EFD) and the internal (IFD) \citep{Sokoloff1998}. Thus the turbulent magnetic cells in the intergalactic space on the line-of-sight cause EDF. $DP_{EFD}$ can be calculated using

%\begin{equation}
%    DP_{EFD} = \exp{(-2\sigma^2\lambda^4)}.
%   \label{DP_EFD}
%\end{equation}

%The factor $\sigma$ is defined as
%The strength of Faraday Dispersion $\sigma$ can be defined as

%\begin{equation}
%    \sigma ^2 = (0.81\langle n_e \rangle \langle B_{turb} %\rangle )^2 Ld/f
%    \label{sigma_square}
%\end{equation}

%where $\langle n_e\rangle$ is the average electron density in cm$^{-3}$, $\langle B_{turb} \rangle$ the average turbulent magnetic field strength  in $\mu$G, $d$ the turbulent cell size in pc, $L$ the path-length in pc and $f$ the filling factor of the turbulent cells causing the depolarisation.

\subsection{Redshift-depolarisation dependence}

Due to the cosmologically significant redshift range we are probing in this analysis, we first explore the effect of differing initial wavelengths. 
%Another aspect that we have to take into account is the difference in the wavelength of the emitted and received radio emission due to redshifting. 
Sources at high redshift, detected in our sample at $\uplambda$20\,cm, emit their actual emission at significantly shorter wavelengths. We can calculate the emitted wavelength $\lambda_i$ in the source's rest-frame for our sample using the standard equation:
\begin{equation}
    \lambda_i = \frac{\lambda_{obs}}{z+1}
\end{equation}
where $\lambda_{obs}$ is the observed wavelength and $z$ is the redshift. For our sample, the emission we receive for sources at $z=1.4$ would originally be emitted at a wavelength of $\lambda\approx8$\,cm.

Since we assume all our sources to be AGN, this shift of wavelength should not have an effect on the initial fractional polarisation. Anyhow, \citet{Conway1977} showed a fractional polarisation of 4\,\% at 21\,cm and of 6\,\% at 6\,cm for a sample of mainly nearby AGN. We attribute this to physical depolarisation effects.

%The contribution of thermal emission to the total power emission increases towards shorter wavelengths in the radio regime. This thermal emission is unpolarised and therefore reduces the overall fractional polarisation. This in turn leads to a lower fractional polarisation for high redshift sources detected at $\lambda = 20\,\text{cm}$, in agreement with our results.
%For a sample of mainly nearby AGN, \citet{Conway1977} showed that this effect is not significant. 
%The authors found a mean fractional polarisation of 4\,\% at 21\,cm and of 6\,\% at 6cm, which we attribute to physical depolarisation effects. 

The shorter wavelength are less affected by IFD and EFD (see equation \ref{eq:DP_EFD}) than the longer wavelengths. This is in contrast to our observed trend, where the higher redshift sources (emitting at a shorter wavelength) are observed to be less polarised than the lower redshift, longer initial wavelength sources.

Therefore, we conclude that we must %only 
account for the different initial wavelengths of emission (due to the location of the sources at different redshifts) and the influence this has on depolarisation.%, but not for the different composition of thermal and synchrotron emission. 

\subsection{Morphology and source environment}

%The morphology and strength of the magnetic field in each source will affect their individual IFD. Assuming all sources to have the same properties regardless of their redshift, we can assume $\sigma$ (see eq. \ref{eq:sigma_square}) to be constant. In this regime, we must only account for the different initial wavelengths.
%Since high redshift sources emit at shorter wavelength in the rest-frame they are less effected by IFD (see eq. \ref{eq:DP_IFD}), and thus exhibit higher fractional polarisations compared to their low redshift counterparts. This is in contrast to our observed trend where the fractional polarisation decreases with redshift.

We now assume the morphology and environments of sources change as a function of redshift, due to the evolution of cosmic structures over time. 
The high redshift sources for which the observed anticorrelation is stronger are mostly identified as strong AGN. We assume them to be mostly FR\,II sources and thus members of galaxy groups  \citep{Ledlow1996}.
%A significant fraction of our sources, due to their FR\,II nature, will be members of galaxy clusters \citep{Ledlow1996}. 
These galaxy groups were less relaxed in the earlier stages of the Universe \citep{2017ApJ...842L..21N}, which can influence their magnetic field strengths and morphologies (and therefore their intrinsic polarisation characteristics). 
%Polarised emission in FR\,II sources is mostly located in their lobes, at termination shocks where the AGN jets become subsonic. These lobes could, due to a denser and more turbulent intracluster medium surrounding them, be less pronounced at higher redshifts, and therefore show smaller polarisation values there.
That these sources have such a high radio luminosity, hints to them being jet dominated. In combination with the less relaxed clusters, this might 
%Additionally this might 
cause a stronger impact of the Laing-Garrington effect \citep{Garrington1988}, which leads to a higher depolarisation of the jet further away from the observer. Thus, for average polarisation values of these sources, we would only be able to detect a smaller fraction of the polarised emission.
%The denser cluster medium could also cause an increased accretion rate and higher AGN activity, accompanied by faster duty cycles. This would lead to differences in the lobe structure of these earlier AGN and bias their properties towards lower overall polarised fractions.

For extended sources we also have to account for beam depolarisation since the projected angular size of sources decrease with increasing redshift up to $z\approx0.5$. For higher redshifts the projected angular size stays nearly constant with increasing redshift \citep{2011A&A...526A...8M} up to a value of $z=1.5$, where it turns around. We do not have sources beyond this limit in our sample. So that the decrease of fractional polarisation for redshifts of $z>0.6$ shown in Fig. \ref{Redshift_FR} cannot be explained by this effect.
%The difference in the appearance of FR\,I and FR\,II sources is often referred to as originating from the different surrounding environments.
%The fact that we see an increase of fractional polarisation for the FR\,I sources for $z<0.6$ and the opposite trend for the FR\,IIs for $z\geq0.6$ indicates that these differences might be as important as the intergalactic magnetic field properties, as discussed in the following subsections.

\subsection{External Faraday Dispersion}
In the previous subsection we considered the sources' direct environment as the primary cause of depolarisation. 
We now want to consider the medium along the line-of-sight as a possible origin of the observed trend. 
Polarised emission of sources further away cross a larger volume of magneto-ionic medium. 
This increases the chance for the (initially polarised) emission to encounter turbulent magnetic cells in intergalactic space along the line-of-sight, which induce depolarisation due to EFD.  
In this case $\sigma$ is no longer a constant for all sources, as the path-length $L$ is dependent on the redshift.
%\begin{equation}
%    L = \frac{\rm c}{{\rm H}_0} \left( z+ \frac{z^2}{2}\right)
%\end{equation}
%where $\rm c$ is the speed of light.
%Assuming $\langle n_e \rangle$ to be in the range $10^{-7} -  10^{-5}\,\text{cm}^{-3}$ \citep{O'Sullivan2020,Vernstrom2019} and $\langle B_{tubr} \rangle = 0.037\,\mu$G \citep{Vernstrom2019}, a turbulent cell size of 10\,kpc and a filling factor of 1, we calculated $DP_{EFD}$ for $\lambda =$ 20 and $\lambda =$ 6\,cm at a range of different redshifts up to $z=2$. 
%For our calculation we assume $\lambda =$ 6\,cm as an upper limit for the depolarisation, to evaluate the strongest significance of the effect. This wavelength corresponds approximatelty to the rest-frame emission of sources with the highest detected redshift in our sample. 
Anyhow, this increasing $L$ does not have a significant effect on the EFD if we assume $\langle B_{tubr} \rangle = 0.037\,\mu$G \citep{Vernstrom2019} being the turbulent magnetic field strength over the whole line of sight.
%Due to the low magnetic field strength and the low electron density, we find $DP_{EFD} \sim 0.9999$ for all combinations of $\lambda$ and $z$ in our sample. 

We therefore conclude that external Faraday Dispersion induced by a turbulent intergalactic magnetic field cannot explain the observed anti-correlation found, assuming a constant electron density and a constant magnetic field strength.

%\subsection{External Faraday Dispersion with evolving universe}

%As we have shown, assuming constant electron densities and magnetic field strengths known from the nearby Universe over the entire line-of-sight, it is not possible to explain the depolarisation of the high redshift sources.

%However, in this calculation we ignored the redshift of the polarised emission over the line-of-sight. By using $\lambda = 20\,\text{cm}$ over the whole line-of-sight we overestimate the EFD, as can be seen by the $\lambda$-dependency in Eq. \ref{eq:DP_EFD}.

%Nonetheless, the approximation we adopted demonstrates that the combined low electron density and low magnetic field strengths we observe in the nearby Universe cannot explain the observed depolarisation.
%This observation is in good agreement with previous works. 
This is in agreement with previous works that assume the electron density as well as the magnetic fields strength to be dependent on z.
For a cosmological significant redshift, however, we cannot assume $n_e$ and $B$ to be constant: rather we expect to find $n_e \propto (1+z)^3$ \citep{O'Sullivan2020, Blasi1999, Pshirokv2016}. 
This increase of the electron density might also lead to an increase of the magnetic fields strength as a function of redshift: $B(z) = B_0[n_e(z)/n_0(z)]$ \citep{O'Sullivan2020}. 
It is also unlikely that, for an increasing magnetic field strength, the turbulence remains constant. This implies that we must also assume a possibly redshift dependent $\langle B_{turb} \rangle$.
An increase of $\langle n_e\rangle$ and $\langle B_{turb}\rangle$ with redshift may induce a higher depolarisation for more distant sources, and thus explain the anti-correlation found between fractional polarisation and redshift.
However, this also means that $L$, $\lambda$, $\langle n_e\rangle$ and $\langle B_{turb}\rangle$ are all interconnected and redshift dependent.
An exploration of these effects, though, would require an extension of Eq. \ref{eq:DP_EFD} and \ref{eq:sigma_square}; this is naturally a considerable task, and we leave such an investigation to future works.

%-----------------------------------------------------------------

\section{Discussion}
\label{sec:discussion}

%We now want to discuss our results in the context of cosmic variance, cosmic evolution and magnetic field morphology. For this we first look into the currently available information on the faint polarised sky and the different observations used to describe it. Secondly, we will highlight the multitude of effects and differences in source characteristics and composition, which can influence the appearance of the faint polarised sky.

\subsection{Observations of the faint polarised sky}
\label{sec:faint_sky}
In Sec. \ref{sec:cosmic_variance} we showed that the different deep field samples show different behaviours of fractional polarisation towards fainter polarised fluxes. One could argue that these different trends were caused by one or more of: instrumental effects, differences in the telescopes used, different software, and differences in data reduction strategies. Countering these possibilities, however, are the two fields analysed by \cite{Hales2014b}, which still show a difference in their trends between each other. These observations were conducted using the same telescope array, software, and reduction strategies, suggesting that other explanations for these polarisation effects should be considered.

Nearly all presented analyses are observations of survey areas spanning up to $15\,\text{deg}^2$, meaning that differences can be caused by sample variance: inhomogenities in the source distribution on-sky caused by differences in large scale structures within a small field. %, which is known as cosmic variance. 
In addition to this, the observed fields on-sky are not randomly selected, but rather are selected specifically due to some special characteristic: the Lockman Hole, for example, shows the lowest HI column density over the entire sky \citep{Lockman1986}. Sources observed in this field are known to have a systematically high median redshift compared to other fields of equivalent depth \citep[$z\sim 0.99$ in][which is similar to our sample's median redshift of $z\sim 0.75$]{Luchsinger2015}.
The ELAIS fields are selected because of their low $100\,\mu$m intensity and their high galactic latitude \citep{Oliver2000}.
The CDF-S has like the Lockman Hole a low HI column density \citep{Giacconi1999}.
All of these preferential field selections may introduce biases in the detected source distributions for these fields, when compared to a statistically `normal' patch of sky of equivalent area. 

\subsection{Spectral index - fractional polarisation dependence}
In Sec. \ref{sec:Spectral_results} we compared the fractional polarisation of steep and flat spectrum sources depending on their total intensity. In contrast to \cite{Tucci2004}, but in agreement with \cite{Mesa2002}, we find no difference in the correlation between spectral index and fractional polarisation for steep and flat spectrum sources. 
In agreement with \citet{Stil2015} steep spectrum sources in our sample are, however, generally more polarised than flat spectrum sources; a difference not observed by either \cite{Mesa2002} nor \cite{Tucci2004}.
It is worth noting, though, that both these previous studies used the spectral index measured between 1.4\,GHz and 5\,GHz, whereas we use the spectral index measured between 1.4\,GHz and 150\,MHz and \cite{Stil2015} use the spectral index measured between 1.4\,GHz and 325\,MHz.
At these low frequencies we must account for additional absorption effects like synchrotron self-absorption and thermal absorption, which lead to lower measured 150\,MHz fluxes and thus a flattening of the spectral index (compared to the spectral indices measured at higher frequencies).
As a result, some of our flat spectrum sources might indeed be defined as steep spectrum sources at higher frequencies. 
This might decrease the difference in fractional polarisation between steep and flat spectrum sources, and reconcile the difference found between our results and those of \cite{Mesa2002,Tucci2004}. 
This explanation is also supported by the fact, that \cite{Stil2015} also observe a higher fractional polarisation for steep spectrum sources than for flat spectrum sources and they use the spectral index towards the MHz-regime.
%found such a huge difference in fractional polarisation between steep and flat spectrum sources, although \citet{Tucci2004} found different trends for the two types of sources. 
%This higher difference might be explained by the different frequencies used for deriving the spectral index. 

It should also be recognised that we are observing the faint polarised radio-sky, while \cite{Tucci2004} and \cite{Mesa2002}  are working with the bright polarised radio-sky. \cite{Stil2015} are probing the range in between, due to their stacking technique. Even in the bright regime they observe a strong difference between steep and flat spectrum sources.
%Unfortunately, though, we are not able to verify the influence of these different observational modes, as we do not have a statistically significant number of bright sources with which to compare to the data from \citet{Mesa2002} and \citet{Tucci2004}.
Therefore, we conclude that the different wavelengths used for the derivation of the spectral indices are causing the different observed median fractional polarisation for steep and flat spectrum sources.
Due to this we can 
%Therefore, due to the different wavelengths used for the derivation of the spectral indices and the different overall brightness of the samples used, we can 
neither confirm nor contradict the different trends for steep and flat spectrum sources observed by \citet{Tucci2004} with our observation.

\subsection{Source counts}
\label{sec:source_counts_discussion}
\begin{figure}
    \centering
    \includegraphics[width = \linewidth]{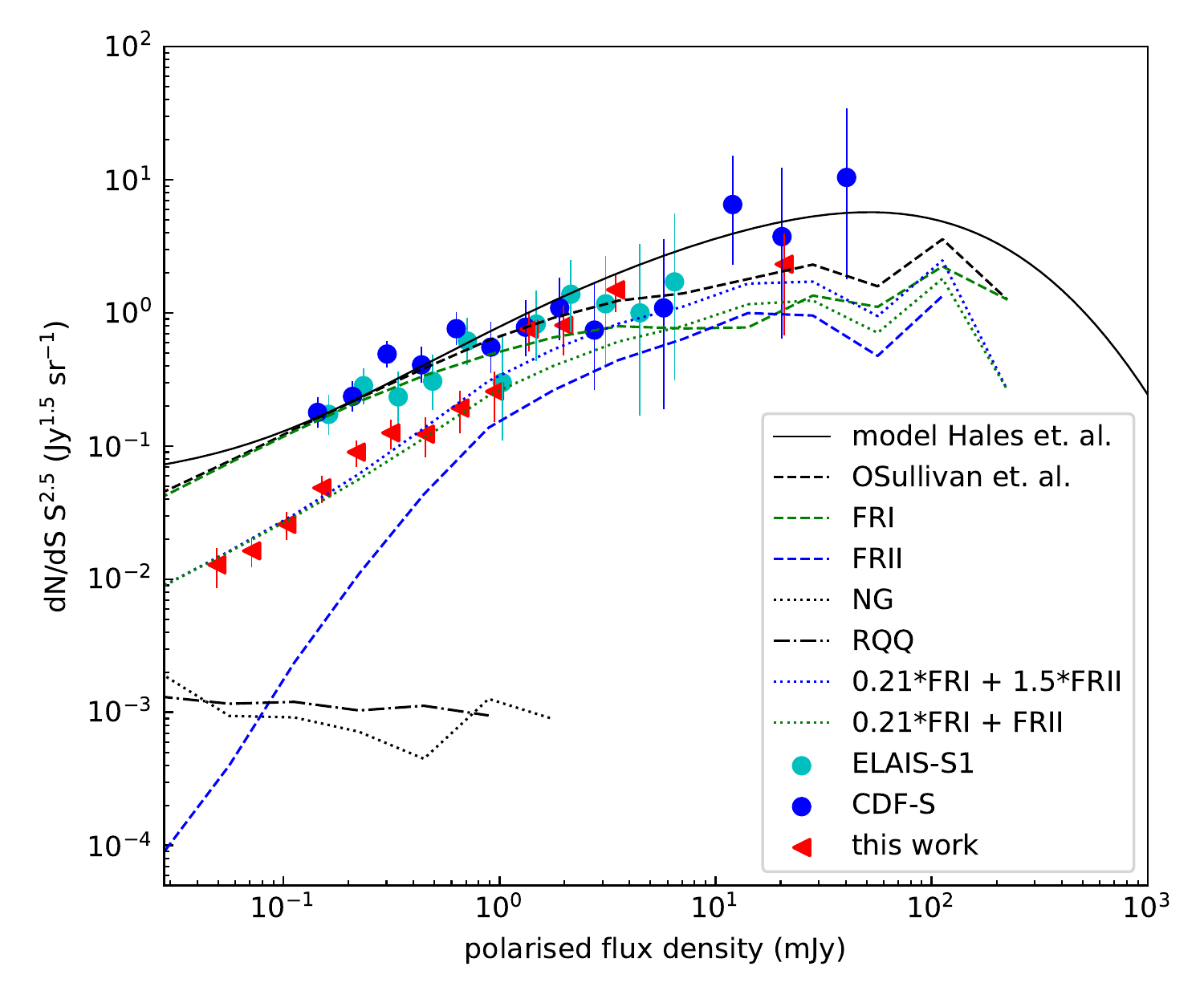}
    \caption{The same plot giving the euclidian normalisied differential source counts like in fig. \ref{fig:diff_source_counts}. The additional curves are defined by the summation of the source counts from \cite{O'Sullivan2008} $a1\cdot \text{FR\,I} + a2 \cdot \text{FR\,II}$, where for the green dotted line $a1 = 0.21$ and $a2 = 1$ and for the blue dotted line $a1 = 0.21$ and $a2 = 1.5$}
    \label{fig:diff_source_counts_model}
\end{figure}

In Sec. \ref{sec:source_counts} we presented the differential Euclidean normalised polarised source counts of our field, compared to the observations from \cite{Hales2014b} and the simulations from \cite{O'Sullivan2008}.
To low flux densities we observed lower source counts than expected from both previous observations and the models.
%This means that we found less faint sources than expected.
Since we observe sources with a high median redshift $z=0.75$ than previous studies, this difference may be driven by a dearth of faint near sources in the Lockman Hole field.
Fig. \ref{fig:diff_source_counts_model} shows the same plot as given in Fig. \ref{fig:diff_source_counts} but with two additional lines. We used the simulations from \citet{O'Sullivan2008}, for the individual source types, to find a fit for our observations. 
The simulations are based on complex functions which we do not want to replicate here, so we are not fitting our data directly. 
Using the summation of the individual source counts as $a1\cdot \text{FR\,I} +a2 \cdot  \text{FR\,II}$ we end up with the green dotted line using $a1 = 0.21$ and $a2 = 1$.
That is, we assume that we observe only 21\,\% of the FR\,I sources compared to \cite{O'Sullivan2008}. Since our sample is dominated by bright sources, we %Since we assume all our sources to be AGN, we
ignore the contribution from NG or RQQ. 
The FR sample from \cite{O'Sullivan2008} goes down to flux denstiys of about 1\,mJy and contains $\sim99.88\,\%$ FR\,I sources. Like discussed in sec. \ref{sec:FR-Classification} we cannot classify our sources following the Fanaroff-Riley-classification, but our cutoff for weaker and stronger AGN was previously used for this purpose \citep{Gendre2008}. We only find $\sim21.43\,\%$ of the 56 sources that we were able to classify to be weak, and thus likely FRI.
%by the FR classification are classified as FR\,I. 
We assume those 56 sources to be representative for our whole sample, suggesting that we are biased towards lower FR\,I source numbers. 
%We only detect an amount of ~21\,\% of the FR\,I sources expected from \cite{O'Sullivan2008}, which is represented by the green dotted line in Fig. \ref{fig:diff_source_counts_model}. 
We note, that the simulations from \cite{O'Sullivan2008} are based on simulations from \cite{Wilman2008} and thus extrapolated from the observed source distributions at 151\,MHz, leading to another unknown parameter for our analysis.

The source density observed by \citet{Hales2014b} is between 16 and 23 sources per $\text{deg}^2$, with a central RMS of 25\,$\mu$Jy/beam. %, while their estimated median redshift is about 0.2. 
We observed 23 polarised sources per $\text{deg}^2$ with a lower central RMS of 7$\,\mu$Jy/beam. 
The differences in source density between the different fields from \cite{Hales2014b}, as well as the possible dearth of FR\,I sources in our observations, can originate from the already discussed differences in source characteristics of different deep fields (see Sec. \ref{sec:cosmic_variance} and \ref{sec:faint_sky}).
This difference might also explain a slight difference we have between our source density and the density predicted by \cite{Rudnick2014} from their deep GOODS-N observations of $35\pm10$ sources per deg$^2$ for a flux regime of about 50$\,\mu$Jy, which is in good agreement with our flux regime, taking into account our detection threshold of 6.25\,$\sigma$.

\begin{figure}
    \centering
    \includegraphics[width = \linewidth]{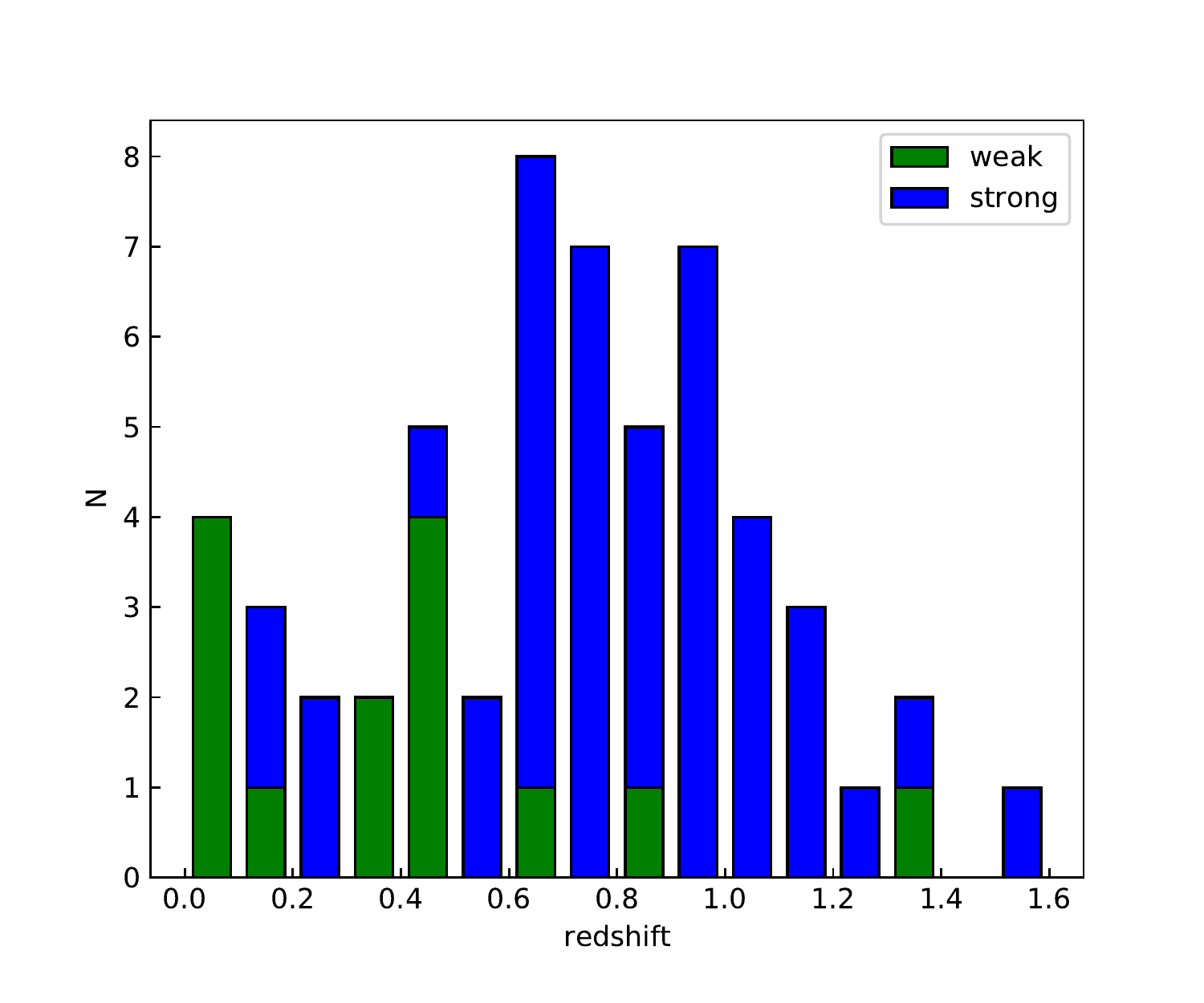}
    \caption{Histogram of the photometric redshifts of the 56 sources classifyed as weak and strong sources.}
    \label{fig:Hist_z}
\end{figure}

In Fig. \ref{fig:Hist_z} we show the redshift distribution of the 56 sources for which we are able to determine redshifts. In agreement with previous observations of the Lockman Hole \citep[e.g.][]{Luchsinger2015, Fotopoulou2012} we observe a dearth of low-redshift sources, up to a redshift of $z \sim 0.4$. 
In Fig. 3 of \cite{Luchsinger2015} a lack of low-$z$ sources is visible, compared to the simulations from \cite{Wilman2008}.
\cite{Luchsinger2015} explained this as being due to the over-resolution of low-z sources, so the emission from near sources is resolved out. 
If this were true, a lower resolution survey of the field should recover this population of low-redshift sources. 
%In Fig. \ref{fig:Hist_z_vergleich}, we compare the redshift distribution of our sample directly to the simulations of \cite{Wilman2008} and to the AGN sample from \cite{Bonaldi2019}.
The resolution of our PI and total intensity map is 15\,\arcsec, while the resolution of \cite{Luchsinger2015} is 4\,\arcsec. Nonetheless, we recover the same source redshift distribution as \cite{Luchsinger2015}. We therefore conclude that this dearth of low-redshift sources compared to the simulation of \cite{Wilman2008} is not due to them being over-resolved and lost from the imaging. 
Further, the observations of \citet{Fotopoulou2012} show the same redshift distribution for X-ray sources and normal optical galaxies.
It is thus possible that the difference between the Lockman Hole observations and the simulations of \citet{Wilman2008} and the observations of \cite{Hales2014a} are driven by a combination of large scale structure variations (i.e. sample variance), which at our estimated distances of up to $z\approx 1.4$ can fill an entire field of the size of the Lockman Hole, and by differences in source properties between small survey areas. Finally, FR\,I sources dominate the observations at low redshift \citep{Ledlow1996}. We can therefore conclude that a possible explanation of the low source counts at low flux densities is a dearth of low-redshift FR\,I sources. 
%This is in good agreement with \cite{Luchsinger2015} who also found a lack of near sources compared to the simulations from \cite{Wilman2008}. 

So far we have only taken low flux densities into account in our discussion.
To also account for the higher flux densities, %, where our source counts are than the green dotted line,
we adjusted the relative values of $a1$ and $a2$ to approximately represent our data. Values of $a1 = 0.21$ and $a2 = 1.5$ were used to generate the blue dotted line in Fig. \ref{fig:diff_source_counts_model}. 
%The simulations are based on complex functions which we do not want to replicate here, so we are not fitting our data directly. 
Since we cannot fit the values to our data, these values are only rough estimates; they are designed only to that a higher fraction of FR\,II sources (i.e. $a2=1.5$) explains higher source counts at high fluxes, while the low fraction of FR\,I sources (i.e. $a1=0.21$) still dominates the low source counts at low fluxes, shown by the green dotted line in Fig. \ref{fig:diff_source_counts_model}.
We note that we therefore cannot exclude other combinations of $a1$ and $a2$ to give similar results.

A higher fraction of FR\,II sources could be due to a large-scale overdensity of sources at high redshift.
This is indicated by the distributions from \cite{Luchsinger2015} (Fig.3) and \cite{Fotopoulou2012} (Fig.11 \& Fig.15), which show peaks at $z \approx 0.6$ and $z \approx 1.5$. 
\cite{Henry2014} also suggest that such an overdense structure is present in the Lockman Hole at  $z=1.71$, which could also contribute to our observations.
Due to the lack of faint sources it might be necessary to distinguish between the bright sources, where our source counts fit the simulations of \cite{O'Sullivan2008}, and faint sources, where the loss of FR\,I sources becomes dominant.

Including NG and RQQ in our fit leads to a flattening of the source counts at around $10^{-1}\,$mJy, which would cause a stronger effect than that which is observed in our data. 
This again fits to our previous assumption: that all detected polarised sources can be classified as AGN.

\begin{figure}
    \centering
    \includegraphics[width = \linewidth]{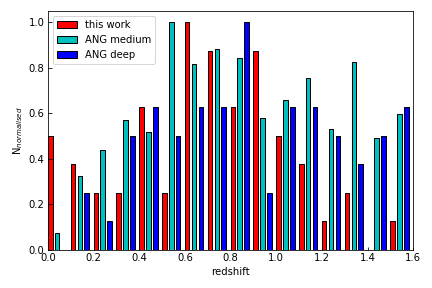}
    \caption{Normed histogram of the photometic redshift of our 56 polarised sources (red), compared to the deep (blue) and medium (cyan) simulation of \cite{Bonaldi2019}, where we used a cutoff of 6.25 times our central RMS. So we only include sources with flux densities, comparable to our sample.  
    We also normed the maximum of each sample to 1 in the histogram, to allow for comparing different sample sizes.}
    \label{fig:Hist_z_vergleich}
\end{figure}

%Since the \cite{Wilman2008} simulation includes as well AGN as star-forming galaxies
For a direct comparison of polarised sources redshift distribution, we compared the redshift distribution of our sample to the simulated AGN sample from \cite{Bonaldi2019}. 
To have an estimate on the effect of local variations of small fields compared to larger ones, we plot the distribution of the deep simulation as well as of the medium simulation from \cite{Bonaldi2019}. 
For comparability, we set a cutoff of 6.25 times our central RMS on the simulated data, to only include sources that would be detectable in our observation.
Fig. \ref{fig:Hist_z_vergleich} shows the normalised histogram of our sources and both samples from \cite{Bonaldi2019}. For low redshifts we detected more sources than expected by the simulation, while we have a loss of high redshift sources. 
Apparently faint sources can either be nearby with low intrinsic (absolute) brightnesses, or be luminous sources located at high redshift. %sources that have a low appeared brightness.
The loss of faint low-$z$ sources was discussed previously, in the context of the loss of near FR\,I sources. 
Fig. \ref{fig:Hist_z_vergleich} hints to a loss of apparently faint sources at high redshift, as another possible reason for our observed source counts.
However, the comparison of the deep and the medium simulation of \cite{Bonaldi2019} shows that, even for the same simulation, there is significant variations caused by the survey area and volume. 
This confirms our conclusion that different fields covering only several square degrees are showing differences in source count statistics of several orders.
This is again in good agreement with our observed differences in the source properties of different, small area fields (see sec. \ref{sec:cosmic_variance} and \ref{sec:faint_sky}).
%Thus we note, that the lower source counts might also just be due to normal scattering.
%Taking into account our high median redshift of 0.75, this means that compared to the sample of \citet{Hales2014b} we must have a lack of low redshift sources, since FR\,I sources dominate the observations at low red shifts \citep{Ledlow1996}. 
%Since only the low flux source counts differ from the model, we conclude that we have a lack of faint low redshift sources.
%This can again be explained by the already discussed effect of cosmic variance. 

\subsection{Impact of cosmic evolution}

In Section \ref{sec:redshift-discussion} we discussed several possible reasons for our observed anti-correlation between fractional polarisation and redshift. 
It is most likely that the origin of this trend can be found in the evolution and structure of the Universe.
It might be that the sources themselves are intrinsically different at different redshifts, due to the evolution of environment with redshift, which also causes different depolarisation effects. 
Finally there are also hints \citep{O'Sullivan2020,Blasi1999,Pshirokv2016} that not only the environment but the cosmic magnetic field changes with an evolving Universe.

%-----------------------------------------------------------------

\section{Summary}

We present a new deep-field analysis of polarised sources in the Lockman Hole at 1.4\,GHz, using a bespoke polarised mosaic with a central RMS of $7\,\mu$Jy/beam. We find 150 polarised sources in an area of ~6.5$\,\text{deg}^2$ out of 1708 total-intensity sources in this field (8.8\,\%). This equates to a polarised source density of 23/deg$^2$.

We were able to explore fainter polarised flux densities than previous studies with comparable areas. All our sources were classified as AGN. This result is in agreement with the findings from \cite{Hales2014b}.

We found mainly faint polarised sources with only a few percent polarisation; the mean fractional polarisation is 5.4\,\%. The highest fractional polarisation found is $\sim21\,\%$. 

A direct morphological classification, following the Fanaroff-Riley classification scheme, was not possible for the majority of our sources in the sample, as most sources are unresolved. Using the absolute radio brightness and the optical B-magnitude as a classification criterion, which was previously used for classifying according to the Fanaroff-Riley classification, we were able to classify 14 sources as weak and 42 sources as strong sources.
%FR\,II sources are more likely to be found in the high redshift universe than in the nearby one, where this is contrary for FR\,I sources. 

When comparing our catalogue to other deep field catalogues from previous work, we found significant differences (both between our dataset and previous work, and between individual previous studies) in median fractional polarisation as a function of total intensity. We attribute these differences  to sample variance, caused by large-scale structures within different small on-sky areas, that cause different source populations to be observed within the various fields.

We present euclidean normalised differential source counts for our field. A comparison to the analysis of previous publications shows hints, that our sample is dominated by FR\,II sources, while most other studies are dominated by FR\,I sources. We found hints for a dearth of low-redshift sources in the Lockman Hole field, taking into account observations at other wavelengths. This is again in agreement with previous work, who showed that the Lockman Hole is underdense at low redshift \citep{Luchsinger2015, Fotopoulou2012}.

We were able to cross identify 56 of our sources with a deep optical catalogue containing photometric redshift information. 
Using this redshift information we found an anti-correlation between median fractional polarisation and redshift. 
This trend is visible for the whole sample. Strong sources alone show the same behaviour, while the same trend is not visible for weak sources.
We discuss several possible explanations for this anti-correlation. We conclude that this trend originates in either: host sources being contained within different environments as a function of redshift, and/or in a significant evolution of the cosmic magnetic field (in both field strength and morphology) over the lifetime of the Universe.

\begin{acknowledgements}

The research at AIRUB is supported through BMBF project grants D-LOFAR IV (FKZ: 05A17PC1) and D-MeerKAT (FKZ: 05A17PC2).
%NHR acknowledges support from the BMBF through projects D-LOFAR IV (FKZ: 05A17PC1) and MeerKAT (FKZ: 05A17PC2).
AHW is supported by an European Research Council Consolidator Grant (No. 770935).
IP acknowledges support from INAF under the SKA/CTA PRIN “FORECaST” and the PRIN MAIN STREAM “SAuROS” projects. We want to thank J. M. Stil for providing the data from his publication.
\end{acknowledgements}

% WARNING%%%%%
  \bibliographystyle{aa} % style aa.bst
  \bibliography{bibliothek} % your references Yourfile.bib

\begin{thebibliography}{70}
\expandafter\ifx\csname natexlab\endcsname\relax\def\natexlab#1{#1}\fi

\bibitem[{{Abolfathi} {et~al.}(2018){Abolfathi}, {Aguado}, {Aguilar}, {Allende
  Prieto}, {Almeida}, {Ananna}, {Anders}, {Anderson}, {Andrews}, {Anguiano},
  {Arag{\'o}n-Salamanca}, {Argudo-Fern{\'a}ndez}, {Armengaud}, {Ata},
  {Aubourg}, {Avila-Reese}, {Badenes}, {Bailey}, {Balland}, {Barger},
  {Barrera-Ballesteros}, {Bartosz}, {Bastien}, {Bates}, {Baumgarten},
  {Bautista}, {Beaton}, {Beers}, {Belfiore}, {Bender}, {Bernardi}, {Bershady},
  {Beutler}, {Bird}, {Bizyaev}, {Blanc}, {Blanton}, {Blomqvist}, {Bolton},
  {Boquien}, {Borissova}, {Bovy}, {Bradna Diaz}, {Brandt}, {Brinkmann},
  {Brownstein}, {Bundy}, {Burgasser}, {Burtin}, {Busca}, {Ca{\~n}as},
  {Cano-D{\'\i}az}, {Cappellari}, {Carrera}, {Casey}, {Cervantes Sodi}, {Chen},
  {Cherinka}, {Chiappini}, {Choi}, {Chojnowski}, {Chuang}, {Chung}, {Clerc},
  {Cohen}, {Comerford}, {Comparat}, {Correa do Nascimento}, {da Costa},
  {Cousinou}, {Covey}, {Crane}, {Cruz-Gonzalez}, {Cunha}, {da Silva Ilha},
  {Damke}, {Darling}, {Davidson}, {Dawson}, {de Icaza Lizaola}, {de la
  Macorra}, {de la Torre}, {De Lee}, {de Sainte Agathe}, {Deconto Machado},
  {Dell'Agli}, {Delubac}, {Diamond-Stanic}, {Donor}, {Downes}, {Drory}, {du Mas
  des Bourboux}, {Duckworth}, {Dwelly}, {Dyer}, {Ebelke}, {Davis Eigenbrot},
  {Eisenstein}, {Elsworth}, {Emsellem}, {Eracleous}, {Erfanianfar},
  {Escoffier}, {Fan}, {Fern{\'a}ndez Alvar}, {Fernandez-Trincado}, {Fernand o
  Cirolini}, {Feuillet}, {Finoguenov}, {Fleming}, {Font-Ribera}, {Freischlad},
  {Frinchaboy}, {Fu}, {G{\'o}mez Maqueo Chew}, {Galbany}, {Garc{\'\i}a
  P{\'e}rez}, {Garcia-Dias}, {Garc{\'\i}a-Hern{\'a}ndez}, {Garma Oehmichen},
  {Gaulme}, {Gelfand }, {Gil-Mar{\'\i}n}, {Gillespie}, {Goddard}, {Gonz{\'a}lez
  Hern{\'a}ndez}, {Gonzalez-Perez}, {Grabowski}, {Green}, {Grier}, {Gueguen},
  {Guo}, {Guy}, {Hagen}, {Hall}, {Harding}, {Hasselquist}, {Hawley}, {Hayes},
  {Hearty}, {Hekker}, {Hernand ez}, {Hernandez Toledo}, {Hogg},
  {Holley-Bockelmann}, {Holtzman}, {Hou}, {Hsieh}, {Hunt}, {Hutchinson},
  {Hwang}, {Jimenez Angel}, {Johnson}, {Jones}, {J{\"o}nsson}, {Jullo}, {Khan},
  {Kinemuchi}, {Kirkby}, {Kirkpatrick}, {Kitaura}, {Knapp}, {Kneib},
  {Kollmeier}, {Lacerna}, {Lane}, {Lang}, {Law}, {Le Goff}, {Lee}, {Li}, {Li},
  {Lian}, {Liang}, {Lima}, {Lin}, {Long}, {Lucatello}, {Lundgren}, {Mackereth},
  {MacLeod}, {Mahadevan}, {Maia}, {Majewski}, {Manchado}, {Maraston},
  {Mariappan}, {Marques-Chaves}, {Masseron}, {Masters}, {McDermid}, {McGreer},
  {Melendez}, {Meneses-Goytia}, {Merloni}, {Merrifield}, {Meszaros}, {Meza},
  {Minchev}, {Minniti}, {Mueller}, {Muller-Sanchez}, {Muna}, {Mu{\~n}oz},
  {Myers}, {Nair}, {Nand ra}, {Ness}, {Newman}, {Nichol}, {Nidever},
  {Nitschelm}, {Noterdaeme}, {O'Connell}, {Oelkers}, {Oravetz}, {Oravetz},
  {Ort{\'\i}z}, {Osorio}, {Pace}, {Padilla}, {Palanque-Delabrouille},
  {Palicio}, {Pan}, {Pan}, {Parikh}, {P{\^a}ris}, {Park}, {Peirani},
  {Pellejero-Ibanez}, {Penny}, {Percival}, {Perez-Fournon}, {Petitjean},
  {Pieri}, {Pinsonneault}, {Pisani}, {Prada}, {Prakash}, {Queiroz}, {Raddick},
  {Raichoor}, {Barboza Rembold}, {Richstein}, {Riffel}, {Riffel}, {Rix},
  {Robin}, {Rodr{\'\i}guez Torres}, {Rom{\'a}n-Z{\'u}{\~n}iga}, {Ross},
  {Rossi}, {Ruan}, {Ruggeri}, {Ruiz}, {Salvato}, {S{\'a}nchez}, {S{\'a}nchez},
  {Sanchez Almeida}, {S{\'a}nchez-Gallego}, {Santana Rojas}, {Santiago},
  {Schiavon}, {Schimoia}, {Schlafly}, {Schlegel}, {Schneider}, {Schuster},
  {Schwope}, {Seo}, {Serenelli}, {Shen}, {Shen}, {Shetrone}, {Shull}, {Silva
  Aguirre}, {Simon}, {Skrutskie}, {Slosar}, {Smethurst}, {Smith}, {Sobeck},
  {Somers}, {Souter}, {Souto}, {Spindler}, {Stark}, {Stassun}, {Steinmetz},
  {Stello}, {Storchi-Bergmann}, {Streblyanska}, {Stringfellow}, {Su{\'a}rez},
  {Sun}, {Szigeti}, {Taghizadeh-Popp}, {Talbot}, {Tang}, {Tao}, {Tayar},
  {Tembe}, {Teske}, {Thakar}, {Thomas}, {Tissera}, {Tojeiro}, {Tremonti},
  {Troup}, {Urry}, {Valenzuela}, {van den Bosch}, {Vargas-Gonz{\'a}lez},
  {Vargas-Maga{\~n}a}, {Vazquez}, {Villanova}, {Vogt}, {Wake}, {Wang},
  {Weaver}, {Weijmans}, {Weinberg}, {Westfall}, {Whelan}, {Wilcots}, {Wild},
  {Williams}, {Wilson}, {Wood-Vasey}, {Wylezalek}, {Xiao}, {Yan}, {Yang},
  {Ybarra}, {Y{\`e}che}, {Zakamska}, {Zamora}, {Zarrouk}, {Zasowski}, {Zhang},
  {Zhao}, {Zhao}, {Zheng}, {Zheng}, {Zhou}, {Zhu}, {Zinn}, \&
  {Zou}}]{Abolfathi2018}
{Abolfathi}, B., {Aguado}, D.~S., {Aguilar}, G., {et~al.} 2018, \apjs, 235, 42

\bibitem[{{Adebahr} {et~al.}(2013){Adebahr}, {Krause}, {Klein},
  {We{\.z}gowiec}, {Bomans}, \& {Dettmar}}]{2013A&A...555A..23A}
{Adebahr}, B., {Krause}, M., {Klein}, U., {et~al.} 2013, \aap, 555, A23

\bibitem[{{Appleton} {et~al.}(2004){Appleton}, {Fadda}, {Marleau}, {Frayer},
  {Helou}, {Condon}, {Choi}, {Yan}, {Lacy}, {Wilson}, {Armus}, {Chapman},
  {Fang}, {Heinrichson}, {Im}, {Jannuzi}, {Storrie-Lombardi}, {Shupe},
  {Soifer}, {Squires}, \& {Teplitz}}]{Appleton2004}
{Appleton}, P.~N., {Fadda}, D.~T., {Marleau}, F.~R., {et~al.} 2004, \apjs, 154,
  147

\bibitem[{{Beck}(2015)}]{2015A&ARv..24....4B}
{Beck}, R. 2015, \aapr, 24, 4

\bibitem[{{Blandford} {et~al.}(2019){Blandford}, {Meier}, \&
  {Readhead}}]{2019ARA&A..57..467B}
{Blandford}, R., {Meier}, D., \& {Readhead}, A. 2019, \araa, 57, 467

\bibitem[{{Blasi} {et~al.}(1999){Blasi}, {Burles}, \& {Olinto}}]{Blasi1999}
{Blasi}, P., {Burles}, S., \& {Olinto}, A.~V. 1999, \apjl, 514, L79

\bibitem[{{Bonaldi} {et~al.}(2019){Bonaldi}, {Bonato}, {Galluzzi}, {Harrison},
  {Massardi}, {Kay}, {De Zotti}, \& {Brown}}]{Bonaldi2019}
{Bonaldi}, A., {Bonato}, M., {Galluzzi}, V., {et~al.} 2019, \mnras, 482, 2

\bibitem[{{Boyle} \& {Terlevich}(1998)}]{1998MNRAS.293L..49B}
{Boyle}, B.~J. \& {Terlevich}, R.~J. 1998, \mnras, 293, L49

\bibitem[{{Brentjens} \& {de Bruyn}(2005)}]{Brentjens2005}
{Brentjens}, M.~A. \& {de Bruyn}, A.~G. 2005, \aap, 441, 1217

\bibitem[{{Condon} {et~al.}(1998){Condon}, {Cotton}, {Greisen}, {Yin},
  {Perley}, {Taylor}, \& {Broderick}}]{Condon1998}
{Condon}, J.~J., {Cotton}, W.~D., {Greisen}, E.~W., {et~al.} 1998, \aj, 115,
  1693

\bibitem[{{Conway} {et~al.}(1977){Conway}, {Burn}, \&
  {Vall{\'e}e}}]{Conway1977}
{Conway}, R.~G., {Burn}, B.~J., \& {Vall{\'e}e}, J.~P. 1977, \aaps, 27, 155

\bibitem[{{Fanaroff} \& {Riley}(1974)}]{Fanaroff1974}
{Fanaroff}, B.~L. \& {Riley}, J.~M. 1974, \mnras, 167, 31P

\bibitem[{{Farnes} {et~al.}(2014){Farnes}, {Gaensler}, \&
  {Carretti}}]{2014ApJS..212...15F}
{Farnes}, J.~S., {Gaensler}, B.~M., \& {Carretti}, E. 2014, \apjs, 212, 15

\bibitem[{{Fazio} {et~al.}(2004){Fazio}, {Hora}, {Allen}, {Ashby}, {Barmby},
  {Deutsch}, {Huang}, {Kleiner}, {Marengo}, {Megeath}, {Melnick}, {Pahre},
  {Patten}, {Polizotti}, {Smith}, {Taylor}, {Wang}, {Willner}, {Hoffmann},
  {Pipher}, {Forrest}, {McMurty}, {McCreight}, {McKelvey}, {McMurray}, {Koch},
  {Moseley}, {Arendt}, {Mentzell}, {Marx}, {Losch}, {Mayman}, {Eichhorn},
  {Krebs}, {Jhabvala}, {Gezari}, {Fixsen}, {Flores}, {Shakoorzadeh}, {Jungo},
  {Hakun}, {Workman}, {Karpati}, {Kichak}, {Whitley}, {Mann}, {Tollestrup},
  {Eisenhardt}, {Stern}, {Gorjian}, {Bhattacharya}, {Carey}, {Nelson},
  {Glaccum}, {Lacy}, {Lowrance}, {Laine}, {Reach}, {Stauffer}, {Surace},
  {Wilson}, {Wright}, {Hoffman}, {Domingo}, \& {Cohen}}]{IRAC}
{Fazio}, G.~G., {Hora}, J.~L., {Allen}, L.~E., {et~al.} 2004, \apjs, 154, 10

\bibitem[{{Fiore} {et~al.}(2012){Fiore}, {Puccetti}, {Grazian}, {Menci},
  {Shankar}, {Santini}, {Piconcelli}, {Koekemoer}, {Fontana}, {Boutsia},
  {Castellano}, {Lamastra}, {Malacaria}, {Feruglio}, {Mathur}, {Miller}, \&
  {Pannella}}]{2012A&A...537A..16F}
{Fiore}, F., {Puccetti}, S., {Grazian}, A., {et~al.} 2012, \aap, 537, A16

\bibitem[{{Fotopoulou} {et~al.}(2012){Fotopoulou}, {Salvato}, {Hasinger},
  {Rovilos}, {Brusa}, {Egami}, {Lutz}, {Burwitz}, {Henry}, {Huang},
  {Rigopoulou}, \& {Vaccari}}]{Fotopoulou2012}
{Fotopoulou}, S., {Salvato}, M., {Hasinger}, G., {et~al.} 2012, \apjs, 198, 1

\bibitem[{{Garrington} {et~al.}(1988){Garrington}, {Leahy}, {Conway}, \&
  {Laing}}]{Garrington1988}
{Garrington}, S.~T., {Leahy}, J.~P., {Conway}, R.~G., \& {Laing}, R.~A. 1988,
  \nat, 331, 147

\bibitem[{{Gendre} \& {Wall}(2008)}]{Gendre2008}
{Gendre}, M.~A. \& {Wall}, J.~V. 2008, \mnras, 390, 819

\bibitem[{{Giacconi} {et~al.}(1999){Giacconi}, {Rosati}, {Norman}, {Burg},
  {Gilmozzi}, {Tarenghi}, \& {Bergeron}}]{Giacconi1999}
{Giacconi}, R., {Rosati}, P., {Norman}, C., {et~al.} 1999, in Highlights in
  X-ray Astronomy, ed. B.~{Aschenbach} \& M.~J. {Freyberg}, Vol. 272, 419

\bibitem[{{Grant} {et~al.}(2010){Grant}, {Taylor}, {Stil}, {Land ecker},
  {Kothes}, {Ransom}, \& {Scott}}]{Grant2010}
{Grant}, J.~K., {Taylor}, A.~R., {Stil}, J.~M., {et~al.} 2010, \apj, 714, 1689

\bibitem[{{Greisen}(1990)}]{AIPS}
{Greisen}, E.~W. 1990, in Acquisition, Processing and Archiving of Astronomical
  Images, 125--142

\bibitem[{{Hales} {et~al.}(2014{\natexlab{a}}){Hales}, {Norris}, {Gaensler}, \&
  {Middelberg}}]{Hales2014b}
{Hales}, C.~A., {Norris}, R.~P., {Gaensler}, B.~M., \& {Middelberg}, E.
  2014{\natexlab{a}}, \mnras, 440, 3113

\bibitem[{{Hales} {et~al.}(2014{\natexlab{b}}){Hales}, {Norris}, {Gaensler},
  {Middelberg}, {Chow}, {Hopkins}, {Huynh}, {Lenc}, \& {Mao}}]{Hales2014a}
{Hales}, C.~A., {Norris}, R.~P., {Gaensler}, B.~M., {et~al.}
  2014{\natexlab{b}}, \mnras, 441, 2555

\bibitem[{{Hammond} {et~al.}(2012){Hammond}, {Robishaw}, \&
  {Gaensler}}]{2012arXiv1209.1438H}
{Hammond}, A.~M., {Robishaw}, T., \& {Gaensler}, B.~M. 2012, arXiv e-prints,
  arXiv:1209.1438

\bibitem[{{Helfand} {et~al.}(2015){Helfand}, {White}, \&
  {Becker}}]{Helfand2015}
{Helfand}, D.~J., {White}, R.~L., \& {Becker}, R.~H. 2015, VizieR Online Data
  Catalog, VIII/92

\bibitem[{{Henry} {et~al.}(2014){Henry}, {Aoki}, {Finoguenov}, {Fotopoulou},
  {Hasinger}, {salvato}, {Suh}, \& {Tanaka}}]{Henry2014}
{Henry}, J.~P., {Aoki}, K., {Finoguenov}, A., {et~al.} 2014, \apj, 780, 58

\bibitem[{{Herrera Ruiz} {et~al.}(2018){Herrera Ruiz}, {Middelberg}, {Deller},
  {Smol{\v{c}}i{\'c}}, {Norris}, {Novak}, {Delvecchio}, {Best}, {Schinnerer},
  {Momjian}, {Dettmar}, {Brisken}, {Koekemoer}, \& {Scoville}}]{Noelia2018}
{Herrera Ruiz}, N., {Middelberg}, E., {Deller}, A., {et~al.} 2018, \aap, 616,
  A128

\bibitem[{{Hopkins} {et~al.}(2003){Hopkins}, {Afonso}, {Chan}, {Cram},
  {Georgakakis}, \& {Mobasher}}]{Hopkins2003}
{Hopkins}, A.~M., {Afonso}, J., {Chan}, B., {et~al.} 2003, \aj, 125, 465

\bibitem[{{Lamee} {et~al.}(2016){Lamee}, {Rudnick}, {Farnes}, {Carretti},
  {Gaensler}, {Haverkorn}, \& {Poppi}}]{2016ApJ...829....5L}
{Lamee}, M., {Rudnick}, L., {Farnes}, J.~S., {et~al.} 2016, \apj, 829, 5

\bibitem[{{Ledlow} \& {Owen}(1996)}]{Ledlow1996}
{Ledlow}, M.~J. \& {Owen}, F.~N. 1996, \aj, 112, 9

\bibitem[{{Lockman} {et~al.}(1986){Lockman}, {Jahoda}, \&
  {McCammon}}]{Lockman1986}
{Lockman}, F.~J., {Jahoda}, K., \& {McCammon}, D. 1986, \apj, 302, 432

\bibitem[{{Lonsdale} {et~al.}(2003{\natexlab{a}}){Lonsdale}, {Smith},
  {Rowan-Robinson}, {Surace}, {Shupe}, {Xu}, {Oliver}, {Padgett}, {Fang},
  {Conrow}, {Franceschini}, {Gautier}, {Griffin}, {Hacking}, {Masci},
  {Morrison}, {O'Linger}, {Owen}, {P{\'e}rez-Fournon}, {Pierre}, {Puetter},
  {Stacey}, {Castro}, {Polletta}, {Farrah}, {Jarrett}, {Frayer}, {Siana},
  {Babbedge}, {Dye}, {Fox}, {Gonzalez-Solares}, {Salaman}, {Berta}, {Condon},
  {Dole}, \& {Serjeant}}]{Lonsdale2003}
{Lonsdale}, C.~J., {Smith}, H.~E., {Rowan-Robinson}, M., {et~al.}
  2003{\natexlab{a}}, \pasp, 115, 897

\bibitem[{{Lonsdale} {et~al.}(2003{\natexlab{b}}){Lonsdale}, {Smith},
  {Rowan-Robinson}, {Surace}, {Shupe}, {Xu}, {Oliver}, {Padgett}, {Fang},
  {Conrow}, {Franceschini}, {Gautier}, {Griffin}, {Hacking}, {Masci},
  {Morrison}, {O'Linger}, {Owen}, {P{\'e}rez-Fournon}, {Pierre}, {Puetter},
  {Stacey}, {Castro}, {Polletta}, {Farrah}, {Jarrett}, {Frayer}, {Siana},
  {Babbedge}, {Dye}, {Fox}, {Gonzalez-Solares}, {Salaman}, {Berta}, {Condon},
  {Dole}, \& {Serjeant}}]{SWIRE}
{Lonsdale}, C.~J., {Smith}, H.~E., {Rowan-Robinson}, M., {et~al.}
  2003{\natexlab{b}}, \pasp, 115, 897

\bibitem[{{Luchsinger} {et~al.}(2015){Luchsinger}, {Lacy}, {Jones}, {Mauduit},
  {Pforr}, {Surace}, {Vaccari}, {Farrah}, {Gonzales-Solares}, {Jarvis},
  {Maraston}, {Marchetti}, {Oliver}, {Afonso}, {Cappozi}, \&
  {Sajina}}]{Luchsinger2015}
{Luchsinger}, K.~M., {Lacy}, M., {Jones}, K.~M., {et~al.} 2015, \aj, 150, 87

\bibitem[{{Mahony} {et~al.}(2016){Mahony}, {Morganti}, {Prandoni}, {van
  Bemmel}, {Shimwell}, {Brienza}, {Best}, {Br{\"u}ggen}, {Calistro Rivera}, {de
  Gasperin}, {Hardcastle}, {Harwood}, {Heald}, {Jarvis}, {Mandal}, {Miley},
  {Retana-Montenegro}, {R{\"o}ttgering}, {Sabater}, {Tasse}, {van Velzen}, {van
  Weeren}, {Williams}, \& {White}}]{Mahony2016}
{Mahony}, E.~K., {Morganti}, R., {Prandoni}, I., {et~al.} 2016, \mnras, 463,
  2997

\bibitem[{{Martin} {et~al.}(2005){Martin}, {Fanson}, {Schiminovich},
  {Morrissey}, {Friedman}, {Barlow}, {Conrow}, {Grange}, {Jelinsky},
  {Milliard}, {Siegmund}, {Bianchi}, {Byun}, {Donas}, {Forster}, {Heckman},
  {Lee}, {Madore}, {Malina}, {Neff}, {Rich}, {Small}, {Surber}, {Szalay},
  {Welsh}, \& {Wyder}}]{Martin2005}
{Martin}, D.~C., {Fanson}, J., {Schiminovich}, D., {et~al.} 2005, \apjl, 619,
  L1

\bibitem[{{McMullin} {et~al.}(2007){McMullin}, {Waters}, {Schiebel}, {Young},
  \& {Golap}}]{CASA}
{McMullin}, J.~P., {Waters}, B., {Schiebel}, D., {Young}, W., \& {Golap}, K.
  2007, in Astronomical Society of the Pacific Conference Series, Vol. 376,
  Astronomical Data Analysis Software and Systems XVI, ed. R.~A. {Shaw},
  F.~{Hill}, \& D.~J. {Bell}, 127

\bibitem[{{Mesa} {et~al.}(2002){Mesa}, {Baccigalupi}, {De Zotti}, {Gregorini},
  {Mack}, {Vigotti}, \& {Klein}}]{Mesa2002}
{Mesa}, D., {Baccigalupi}, C., {De Zotti}, G., {et~al.} 2002, \aap, 396, 463

\bibitem[{{Middelberg} {et~al.}(2011){Middelberg}, {Norris}, {Hales},
  {Seymour}, {Johnston-Hollitt}, {Huynh}, {Lenc}, \&
  {Mao}}]{2011A&A...526A...8M}
{Middelberg}, E., {Norris}, R.~P., {Hales}, C.~A., {et~al.} 2011, \aap, 526, A8

\bibitem[{{Mignano} {et~al.}(2008){Mignano}, {Prandoni}, {Gregorini}, {Parma},
  {de Ruiter}, {Wieringa}, {Vettolani}, \& {Ekers}}]{2008A&A...477..459M}
{Mignano}, A., {Prandoni}, I., {Gregorini}, L., {et~al.} 2008, \aap, 477, 459

\bibitem[{{Mingo} {et~al.}(2019){Mingo}, {Croston}, {Hardcastle}, {Best},
  {Duncan}, {Morganti}, {Rottgering}, {Sabater}, {Shimwell}, {Williams},
  {Brienza}, {Gurkan}, {Mahatma}, {Morabito}, {Prandoni}, {Bondi}, {Ineson}, \&
  {Mooney}}]{Mingo2019}
{Mingo}, B., {Croston}, J.~H., {Hardcastle}, M.~J., {et~al.} 2019, \mnras, 488,
  2701

\bibitem[{{Mohan} \& {Rafferty}(2015)}]{pyBDSF}
{Mohan}, N. \& {Rafferty}, D. 2015, {PyBDSF: Python Blob Detection and Source
  Finder}

\bibitem[{{Noble} {et~al.}(2017){Noble}, {McDonald}, {Muzzin}, {Nantais},
  {Rudnick}, {van Kampen}, {Webb}, {Wilson}, {Yee}, {Boone}, {Cooper},
  {DeGroot}, {Delahaye}, {Demarco}, {Foltz}, {Hayden}, {Lidman},
  {Manilla-Robles}, \& {Perlmutter}}]{2017ApJ...842L..21N}
{Noble}, A.~G., {McDonald}, M., {Muzzin}, A., {et~al.} 2017, \apjl, 842, L21

\bibitem[{{Offringa} {et~al.}(2010){Offringa}, {de Bruyn}, {Biehl}, {Zaroubi},
  {Bernardi}, \& {Pandey}}]{Offringa2010}
{Offringa}, A.~R., {de Bruyn}, A.~G., {Biehl}, M., {et~al.} 2010, \mnras, 405,
  155

\bibitem[{{Oliver} {et~al.}(2000){Oliver}, {Rowan-Robinson}, {Alexander},
  {Almaini}, {Balcells}, {Baker}, {Barcons}, {Barden}, {Bellas-Velidis},
  {Cabrera-Guerra}, {Carballo}, {Cesarsky}, {Ciliegi}, {Clements}, {Crockett},
  {Danese}, {Dapergolas}, {Drolias}, {Eaton}, {Efstathiou}, {Egami}, {Elbaz},
  {Fadda}, {Fox}, {Franceschini}, {Genzel}, {Goldschmidt}, {Graham},
  {Gonzalez-Serrano}, {Gonzalez-Solares}, {Granato}, {Gruppioni},
  {Herbstmeier}, {H{\'e}raudeau}, {Joshi}, {Kontizas}, {Kontizas},
  {Kotilainen}, {Kunze}, {La Franca}, {Lari}, {Lawrence}, {Lemke},
  {Linden-V{\o}rnle}, {Mann}, {M{\'a}rquez}, {Masegosa}, {Mattila}, {McMahon},
  {Miley}, {Missoulis}, {Mobasher}, {Morel}, {N{\o}rgaard-Nielsen}, {Omont},
  {Papadopoulos}, {Perez-Fournon}, {Puget}, {Rigopoulou}, {Rocca-Volmerange},
  {Serjeant}, {Silva}, {Sumner}, {Surace}, {Vaisanen}, {van der Werf}, {Verma},
  {Vigroux}, {Villar-Martin}, \& {Willott}}]{Oliver2000}
{Oliver}, S., {Rowan-Robinson}, M., {Alexander}, D.~M., {et~al.} 2000, \mnras,
  316, 749

\bibitem[{{Oliver} {et~al.}(2012){Oliver}, {Bock}, {Altieri}, {Amblard},
  {Arumugam}, {Aussel}, {Babbedge}, {Beelen}, {B{\'e}thermin}, {Blain},
  {Boselli}, {Bridge}, {Brisbin}, {Buat}, {Burgarella},
  {Castro-Rodr{\'\i}guez}, {Cava}, {Chanial}, {Cirasuolo}, {Clements},
  {Conley}, {Conversi}, {Cooray}, {Dowell}, {Dubois}, {Dwek}, {Dye}, {Eales},
  {Elbaz}, {Farrah}, {Feltre}, {Ferrero}, {Fiolet}, {Fox}, {Franceschini},
  {Gear}, {Giovannoli}, {Glenn}, {Gong}, {Gonz{\'a}lez Solares}, {Griffin},
  {Halpern}, {Harwit}, {Hatziminaoglou}, {Heinis}, {Hurley}, {Hwang}, {Hyde},
  {Ibar}, {Ilbert}, {Isaak}, {Ivison}, {Lagache}, {Le Floc'h}, {Levenson},
  {Faro}, {Lu}, {Madden}, {Maffei}, {Magdis}, {Mainetti}, {Marchetti},
  {Marsden}, {Marshall}, {Mortier}, {Nguyen}, {O'Halloran}, {Omont}, {Page},
  {Panuzzo}, {Papageorgiou}, {Patel}, {Pearson}, {P{\'e}rez-Fournon}, {Pohlen},
  {Rawlings}, {Raymond}, {Rigopoulou}, {Riguccini}, {Rizzo}, {Rodighiero},
  {Roseboom}, {Rowan-Robinson}, {S{\'a}nchez Portal}, {Schulz}, {Scott},
  {Seymour}, {Shupe}, {Smith}, {Stevens}, {Symeonidis}, {Trichas}, {Tugwell},
  {Vaccari}, {Valtchanov}, {Vieira}, {Viero}, {Vigroux}, {Wang}, {Ward},
  {Wardlow}, {Wright}, {Xu}, \& {Zemcov}}]{Oliver2012}
{Oliver}, S.~J., {Bock}, J., {Altieri}, B., {et~al.} 2012, \mnras, 424, 1614

\bibitem[{{O'Sullivan} {et~al.}(2008){O'Sullivan}, {Stil}, {Taylor}, {Ricci},
  {Grant}, \& {Shorten}}]{O'Sullivan2008}
{O'Sullivan}, S., {Stil}, J., {Taylor}, A.~R., {et~al.} 2008, in The role of
  VLBI in the Golden Age for Radio Astronomy, Vol.~9, 107

\bibitem[{{O'Sullivan} {et~al.}(2020){O'Sullivan}, {Br{\"u}ggen}, {Vazza},
  {Carretti}, {Locatelli}, {Stuardi}, {Vacca}, {Vernstrom}, {Heald},
  {Horellou}, {Shimwell}, {Hardcastle}, {Tasse}, \&
  {R{\"o}ttgering}}]{O'Sullivan2020}
{O'Sullivan}, S.~P., {Br{\"u}ggen}, M., {Vazza}, F., {et~al.} 2020, arXiv
  e-prints, arXiv:2002.06924

\bibitem[{{Perley} \& {Butler}(2017)}]{Perley2017}
{Perley}, R.~A. \& {Butler}, B.~J. 2017, \apjs, 230, 7

\bibitem[{{Prandoni} {et~al.}(2018){Prandoni}, {Guglielmino}, {Morganti},
  {Vaccari}, {Maini}, {R{\"o}ttgering}, {Jarvis}, \& {Garrett}}]{LH_Total}
{Prandoni}, I., {Guglielmino}, G., {Morganti}, R., {et~al.} 2018, \mnras, 481,
  4548

\bibitem[{{Pshirkov} {et~al.}(2016){Pshirkov}, {Tinyakov}, \&
  {Urban}}]{Pshirokv2016}
{Pshirkov}, M.~S., {Tinyakov}, P.~G., \& {Urban}, F.~R. 2016, \prl, 116, 191302

\bibitem[{{Rieke} {et~al.}(2004){Rieke}, {Young}, {Engelbracht}, {Kelly},
  {Low}, {Haller}, {Beeman}, {Gordon}, {Stansberry}, {Misselt}, {Cadien},
  {Morrison}, {Rivlis}, {Latter}, {Noriega-Crespo}, {Padgett}, {Stapelfeldt},
  {Hines}, {Egami}, {Muzerolle}, {Alonso-Herrero}, {Blaylock}, {Dole}, {Hinz},
  {Le Floc'h}, {Papovich}, {P{\'e}rez-Gonz{\'a}lez}, {Smith}, {Su}, {Bennett},
  {Frayer}, {Henderson}, {Lu}, {Masci}, {Pesenson}, {Rebull}, {Rho}, {Keene},
  {Stolovy}, {Wachter}, {Wheaton}, {Werner}, \& {Richards}}]{MIPS}
{Rieke}, G.~H., {Young}, E.~T., {Engelbracht}, C.~W., {et~al.} 2004, \apjs,
  154, 25

\bibitem[{{Rudnick} \& {Owen}(2014)}]{Rudnick2014}
{Rudnick}, L. \& {Owen}, F.~N. 2014, \apj, 785, 45

\bibitem[{{Saikia} \& {Salter}(1988)}]{1988ARA&A..26...93S}
{Saikia}, D.~J. \& {Salter}, C.~J. 1988, \araa, 26, 93

\bibitem[{{Sajina} {et~al.}(2005){Sajina}, {Lacy}, \& {Scott}}]{Sajina2005}
{Sajina}, A., {Lacy}, M., \& {Scott}, D. 2005, \apj, 621, 256

\bibitem[{{Sault} {et~al.}(1995){Sault}, {Teuben}, \& {Wright}}]{MIRIAD}
{Sault}, R.~J., {Teuben}, P.~J., \& {Wright}, M.~C.~H. 1995, in Astronomical
  Society of the Pacific Conference Series, Vol.~77, Astronomical Data Analysis
  Software and Systems IV, ed. R.~A. {Shaw}, H.~E. {Payne}, \& J.~J.~E.
  {Hayes}, 433

\bibitem[{{Sokoloff} {et~al.}(1998){Sokoloff}, {Bykov}, {Shukurov},
  {Berkhuijsen}, {Beck}, \& {Poezd}}]{Sokoloff1998}
{Sokoloff}, D.~D., {Bykov}, A.~A., {Shukurov}, A., {et~al.} 1998, \mnras, 299,
  189

\bibitem[{{Stil} \& {Keller}(2015)}]{Stil2015}
{Stil}, J. \& {Keller}, B. 2015, in Advancing Astrophysics with the Square
  Kilometre Array (AASKA14), 112

\bibitem[{{Stil} {et~al.}(2014){Stil}, {Keller}, {George}, \&
  {Taylor}}]{Stil2014}
{Stil}, J.~M., {Keller}, B.~W., {George}, S.~J., \& {Taylor}, A.~R. 2014, \apj,
  787, 99

\bibitem[{{Subrahmanyan} {et~al.}(2010){Subrahmanyan}, {Ekers}, {Saripalli}, \&
  {Sadler}}]{Subrahmanyan2010}
{Subrahmanyan}, R., {Ekers}, R.~D., {Saripalli}, L., \& {Sadler}, E.~M. 2010,
  \mnras, 402, 2792

\bibitem[{{Taniguchi}(2004)}]{2004PThPS.155..202T}
{Taniguchi}, Y. 2004, Progress of Theoretical Physics Supplement, 155, 202

\bibitem[{{Taylor} {et~al.}(2007){Taylor}, {Stil}, {Grant}, {Land ecker},
  {Kothes}, {Reid}, {Gray}, {Scott}, {Martin}, {Boothroyd}, {Joncas},
  {Lockman}, {English}, {Sajina}, \& {Bond}}]{Taylor2007}
{Taylor}, A.~R., {Stil}, J.~M., {Grant}, J.~K., {et~al.} 2007, \apj, 666, 201

\bibitem[{{Tucci} {et~al.}(2004){Tucci}, {Mart{\'\i}nez-Gonz{\'a}lez},
  {Toffolatti}, {Gonz{\'a}lez-Nuevo}, \& {De Zotti}}]{Tucci2004}
{Tucci}, M., {Mart{\'\i}nez-Gonz{\'a}lez}, E., {Toffolatti}, L.,
  {Gonz{\'a}lez-Nuevo}, J., \& {De Zotti}, G. 2004, \mnras, 349, 1267

\bibitem[{{Tudorica} {et~al.}(2017){Tudorica}, {Hildebrandt}, {Tewes},
  {Hoekstra}, {Morrison}, {Muzzin}, {Wilson}, {Yee}, {Lidman}, {Hicks},
  {Nantais}, {Erben}, {van der Burg}, \& {Demarco}}]{Hendrik}
{Tudorica}, A., {Hildebrandt}, H., {Tewes}, M., {et~al.} 2017, \aap, 608, A141

\bibitem[{{Vernstrom} {et~al.}(2019){Vernstrom}, {Gaensler}, {Rudnick}, \&
  {Andernach}}]{Vernstrom2019}
{Vernstrom}, T., {Gaensler}, B.~M., {Rudnick}, L., \& {Andernach}, H. 2019,
  \apj, 878, 92

\bibitem[{{Vernstrom} {et~al.}(2018){Vernstrom}, {Gaensler}, {Vacca}, {Farnes},
  {Haverkorn}, \& {O'Sullivan}}]{2018MNRAS.475.1736V}
{Vernstrom}, T., {Gaensler}, B.~M., {Vacca}, V., {et~al.} 2018, \mnras, 475,
  1736

\bibitem[{{Wall}(1998)}]{1998ASSL..226..129W}
{Wall}, J.~V. 1998, Astrophysics and Space Science Library, Vol. 226,
  {Cosmological Inference from New Radio Surveys}, ed. M.~N. {Bremer},
  N.~{Jackson}, \& I.~{Perez-Fournon}, 129

\bibitem[{{Wall} {et~al.}(2005){Wall}, {Jackson}, {Shaver}, {Hook}, \&
  {Kellermann}}]{2005A&A...434..133W}
{Wall}, J.~V., {Jackson}, C.~A., {Shaver}, P.~A., {Hook}, I.~M., \&
  {Kellermann}, K.~I. 2005, \aap, 434, 133

\bibitem[{{Werner} {et~al.}(2004){Werner}, {Roellig}, {Low}, {Rieke}, {Rieke},
  {Hoffmann}, {Young}, {Houck}, {Brandl}, {Fazio}, {Hora}, {Gehrz}, {Helou},
  {Soifer}, {Stauffer}, {Keene}, {Eisenhardt}, {Gallagher}, {Gautier}, {Irace},
  {Lawrence}, {Simmons}, {Van Cleve}, {Jura}, {Wright}, \&
  {Cruikshank}}]{Werner2004}
{Werner}, M.~W., {Roellig}, T.~L., {Low}, F.~J., {et~al.} 2004, \apjs, 154, 1

\bibitem[{{Wilman} {et~al.}(2008){Wilman}, {Miller}, {Jarvis}, {Mauch},
  {Levrier}, {Abdalla}, {Rawlings}, {Kl{\"o}ckner}, {Obreschkow}, {Olteanu}, \&
  {Young}}]{Wilman2008}
{Wilman}, R.~J., {Miller}, L., {Jarvis}, M.~J., {et~al.} 2008, \mnras, 388,
  1335

\end{thebibliography}

\begin{appendix}
\section{Additional tables}
\label{A:Catalogue}

\begin{table}
\centering
\caption{Pointing centres}
\label{T:Pointings}
\begin{tabular}{c|c|c}
Number & R.A. (J2000) & Dec (J2000)\\\hline
1 &10:56:23.03 &+57.30.00.00 \\ \hline
2 &10:56:23.03 &+57.55.00.01 \\ \hline
3 &10:56:23.03 &+58.20.00.02\\ \hline
4 &10:56:23.03 &+58.45.00.00\\ \hline
5 &10:53:38.06 &+57.17.29.97\\ \hline
6 &10:53:38.06 &+57.42.29.98\\ \hline
7 &10:53:38.06 &+58.07.30.00\\ \hline
8 &10:53:38.06 &+58.32.30.01\\ \hline
9 &10:50:53.10 &+57.30.00.00\\ \hline
10 &10:50:53.10 &+57.55.00.01\\ \hline
11 &10:50:53.10 &+58.20.00.02\\ \hline
12 &10:50:53.10 &+58.45.00.00 \\ \hline
13 &10:48:08.13 &+57.17.29.97\\ \hline
14 &10:48:08.13 &+57.42.29.98\\ \hline
15 &10:48:08.13 &+58.07.30.00\\ \hline
16 &10:48:08.13 &+58.32.30.01\\ \hline
\end{tabular}
\end{table}

\begin{table}[]
\centering
\caption{Overview of deep field observation from literature.}
\label{A:overview}
\begin{tabular}{l|c|c|c|c|c}
\hline
%\begin{tabular}[c]{@{}l@{}}Field\\ Author\end{tabular}
Field & area  & N$_{I}$ & RMS$_{I}$ & N$_{PI}$ & RMS$_{PI}$ \\ 
 & deg$^2$ &        & $\mu$Jy/beam  &        & $\mu$Jy/beam  \\
 \hline
ELAIS-N1 (1) & 7.43  & 786    & 80          & 83     & 78          \\
\hline
ELAIS-N1 (2) & 15.16 & 958    & 55          & 136    & 45          \\ 
\hline
ATLBS (3) & 8.42 & 1094    &        80   &     &           \\ 
\hline
ELAIS-S1 (4)& 2.766 & 1051   &    30         & 45     &       25      \\
\hline
CDF-S (4) & 3.626 & 1170   &   30          & 85     &  25           \\
\hline
GOODS-N (5)& 0.308 & 496    &    2.4         & 13     &          \\
\hline
LH (6) & 6.5   & 1708   & 30          & 150    & 7\\
\hline
\end{tabular}
\tablefoot{All values are taken from the related papers. The first column gives the observed Field and the related paper. The second column shows the area. Columns three and four give the amount of sources detected in total intensity and the RMS of the image. The fifth and sixth columns give the correspondent values for the polarised sources. For some publications these values can not easily be given, we refer the reader to the original papers.}
\tablebib{(1) \citet{Taylor2007}; (2) \citet{Grant2010}; (3) \citet{Subrahmanyan2010}; (4) \citet{Hales2014b}; (5) \citet{Rudnick2014}, (6) this work}
\end{table}

\begin{table}[]
\centering
\caption{Total radio brightness bins, numbers and median fractional polarisation, for the 56 sources with known photometric redshift.}
\label{A:bins_absolutbrightness}
\begin{tabular}{l|c|c|c|c}
\begin{tabular}[c]{@{}l@{}}bin\\ log(P)\\ W $Hz^{-1}$ \end{tabular} & N$_{weak}$ & $\Pi^{med}_{weak}$  & N$_{strong}$ & $\Pi^{med}_{strong}$  \\
\hline
22-23  & 4 & 3.98 $\pm$ 1.99  & 0  &  \\
\hline
23-24   & 3 & 6.67 $\pm$ 3.85  & 2        & 6.09 $\pm$ 4.31 \\
\hline
24-25  & 7 & 10.58 $\pm$ 4.00 & 11 & 4.24 $\pm$ 1.28   \\
\hline
25-26   & 0 &  & 23 & 2.39 $\pm$ 0.50  \\
\hline
26-27  & 0 & & 6 & 5.38 $\pm$ 2.20  
\end{tabular}
\end{table}

\begin{table}[]
    \centering
    \caption{Euclidean normalised polarised differential source counts, from fig. \ref{fig:diff_source_counts}.}
    \label{tab:source_counts}
    \begin{tabular}{c|c|c|c|c}
    \hline
    $\Delta$PI & $PI_{mean}$ & N & N$_{eff}$ & dN/dS $S^{2.5}$  \\
    mJy & mJy & & sr$^{-1}$ & Jy$^{1.5}$ sr$^{-1}$ \\
    \hline
    0.041 - 0.059  & 0.049 &  9  & 13451.9 & 0.012 \\ \hline
    0.059 - 0.086  & 0.071 &  16 & 10139.5 & 0.016 \\ \hline
    0.086 - 0.125  & 0.103 &  18 & 9004.3  & 0.025 \\ \hline
    0.125 - 0.181  & 0.150 &  19 & 9504.0  & 0.047 \\ \hline
    0.181 - 0.261  & 0.217 &  20 & 10004.2 & 0.087 \\ \hline
    0.261 - 0.377  & 0.313 &  16 & 8003.3  & 0.120 \\ \hline
    0.377 - 0.546  & 0.453 &  9  & 4501.8  & 0.116 \\ \hline
    0.546 - 0.789  & 0.656 &  8  & 4001.6  & 0.181 \\ \hline
    0.789 - 1.14   & 0.948 &  6  & 3001.2  & 0.236 \\ \hline
    1.14 - 1.65    & 1.37  &  10 & 5002.1  & 0.683 \\ \hline
    1.65 - 2.38    & 1.98  &  6  & 3001.2  & 0.718 \\ \hline
    2.38 - 4.98    & 3.44  &  10 & 5002.1  & 1.337 \\ \hline
    4.98 - 10.4    & 7.19  &  0  & 0       & 0 \\ \hline
    10.4 - 41.4    & 20.0  &  2  & 1000.4  & 2.001 \\ \hline
    41.4 - 82.6    & 50.8  &  0  & 0       & 0 \\ \hline
    \end{tabular}
    \tablefoot{The first column gives the bin borders of the polarized flux density, the second the geometric mean of the bin. Column three gives the amount of sources in the given flux bin, while column four gives the effective amount of sources per steradian, corrected for the effective area of the mosaic at the given flux levels. Column five gives the euclidean normalised polarised differential source counts.}
\end{table}

Table \ref{tab:catalogue_beispiel} is a short example of our catalogue. The columns are as follows.

\begin{description}
\item \textit{Column 1} Source name/number; Additional Letters A,B,... indicate the individual components of one specific source. 
\item \textit{Column 2,3,4,5} 1.4\,GHz position of the polarised source with their uncertainties
\item \textit{Column 6,7} integrated polarised flux density and its uncertainty
\item \textit{Column 8,9} integrated total flux density and its uncertainty (only given for sources not for individual components)
\item \textit{Column 10,11} fractional polarisation and its uncertainty
\item \textit{Column 12} S-Code from pyBDSF; for components and single-component sources: ``S'': the source is a single-gaussian source, ``M'': a multi-gaussian source, ``G'' a single-gaussian component in a multi component source; For multi component sources: ``Sum'' for sources identified by pyBDSF, ``Com'' for sources identified by visual inspection and/or comparison with other wavelength. The components of sources identified by hand, have the code ``Com'' attached.
\item \textit{Column 13} flag indicating if the source is not found in the total intensity image and thus taken from \cite{LH_Total} with their source ID.
%\item \textit{Column 11} WSRT crossmatch from \cite{Mahony2016} used for spectral Index
\item \textit{Column 14} spectral index from \cite{Mahony2016}
\item \textit{Column 15} photometric redshift taken from \cite{Hendrik}
\item \textit{Column 16,17} photometric redshift min and max values from \cite{Hendrik}
\item \textit{Column 18} absolute B-magnitude taken from \cite{Hendrik}
\item \textit{Column 19} cross-matched SWIRE \citep{SWIRE} source ID
\end{description}

\begin{sidewaystable}[hbtp]
    \centering
    \caption{Short example of the Lockman Hole source and component catalogue. For column description see Appendix \ref{A:Catalogue} text.}
    \begin{scriptsize}
    \begin{tabular}{|l|r|r|r|r|r|r|r|r|r|r|l|l|r|r|r|r|r|l|}
    \hline
    (1) & (2) & (3) & (4) & (5) & (6) & (7) & (8) & (9) & (10) & (11) & (12) & (13) & (14) & (15) & (16) & (17) & (18) & (19) \\
    \hline
    Isl\_ID & RA & RA$_{\text{err}}$ & DEC & DEC$_{\text{err}}$ & PI & PI$_{\text{err}}$ & I & I$_{\text{err}}$ & fracpol & fracpol$_{\text{err}}$ & S\_Code & Flag & Spec$_\text{Index}$ & Z\_B & Z\_B$_\text{min}$ & Z\_B$_\text{max}$ & MAG\_B & SWIRE\_ID \\
    &&&&& mJy & mJy & mJy & mJy & \% & \% & & & & & & & mag & \\
    \hline
    0 & 165.288515 & 8.44E-5 & 57.45604 & 7.91E-5 & 0.161 & 0.0155 & 4.2098 & 0.0547 & 3.83 & 0.37 & S & None & -1.1 & 0.64 & 0.425 & 0.855 & -20.06 & SWIRE3\_J110109.01+572723.3\\
  1 & 165.271963 & 6.21E-5 & 58.270929 & 5.3E-5 & 0.2991 & 0.018 & 5.8448 & 0.0884 & 5.12 & 0.32 & S & None & -1.1 &  &  &  &  & \\
  2 & 165.30147 & 1.87E-4 & 58.742242 & 2.62E-4 & 0.0718 & 0.0166 & 2.8109 & 0.0711 & 2.56 & 0.59 & S & None & -0.5 &  &  &  &  & \\
  3 & 165.199669 & 2.95E-4 & 58.342442 & 1.58E-4 & 0.2188 & 0.0308 & 10.5454 & 0.0962 & 2.07 & 0.29 & S & None & -0.7 &  &  &  &  & \\
  4 & 165.1137 & 8.14E-5 & 57.639642 & 8.89E-5 & 0.1294 & 0.0142 & 3.0503 & 0.0649 & 4.24 & 0.47 & S & None & -0.8 & 0.74 & 0.512 & 1.36 & -19.73 & SWIRE3\_J110026.72+573823.4\\
  5 & 164.947095 & 6.22E-5 & 57.147552 & 7.95E-5 & 0.2173 & 0.0169 & 2.8411 & 0.0554 & 7.65 & 0.61 & S & None & -0.6 &  &  &  &  & \\
  6 & 164.951329 & 1.7E-5 & 57.354765 & 1.89E-5 & 1.0007 & 0.0185 & 24.0227 & 0.1513 & 4.17 & 0.08 & S & None & -0.7 &  &  &  &  & \\
  7 & 164.90805 & 5.16E-5 & 57.448542 & 4.9E-5 & 0.2657 & 0.0156 & 10.7917 & 0.0817 & 2.46 & 0.15 & S & None & -0.9 & 0.98 & 0.72 & 1.65 & -20.77 & SWIRE3\_J105937.95+572652.0\\
  8 & 165.001375 & 7.66E-5 & 59.240997 & 1.05E-4 & 0.1902 & 0.0172 & 1.4263 & 0.0592 & 13.34 & 1.33 & S & None & -0.6 &  &  &  &  & \\
  9 & 164.885708 & 1.69E-5 & 57.761531 & 1.13E-5 & 1.387 & 0.0184 & 48.0113 & 0.2266 & 2.89 & 0.04 & S & None & -0.8 &  &  &  &  & \\
  10 & 164.811838 & 1.16E-5 & 57.247039 & 1.54E-5 & 2.3796 & 0.0251 & 63.712 & 0.2992 & 3.73 & 0.04 & S & None & -1.2 &  &  &  &  & SWIRE3\_J105915.01+571446.9\\
  11 & 164.770049 & 1.81E-5 & 57.505352 & 2.38E-5 & 1.0745 & 0.0201 & 86.7387 & 0.2595 & 1.24 & 0.02 & S & None & -0.8 &  &  &  &  & SWIRE3\_J105904.79+573018.5\\
  12 & 164.835272 & 1.64E-4 & 59.151311 & 9.78E-5 & 0.3909 & 0.0287 & 4.8179 & 0.1443 & 8.11 & 0.64 & S & None & -1.0 &  &  &  &  & \\
  13 & 164.739858 & 1.76E-4 & 57.717832 & 7.84E-5 & 0.0989 & 0.0134 & 4.0755 & 0.1635 & 2.43 & 0.34 & S & None & -0.7 &  &  &  &  & \\
  14 & 164.771415 & 2.55E-5 & 58.361627 & 1.74E-5 & 1.8051 & 0.0327 & 14.4153 & 0.3449 & 12.52 & 0.38 & M & None & -0.9 &  &  &  &  & \\
  15 & 164.813143 & 4.35E-5 & 59.169122 & 3.77E-5 & 0.3759 & 0.0167 & 3.7584 & 0.0935 & 10.0 & 0.51 & S & None & -0.7 &  &  &  &  & \\
  \hline
    \end{tabular}
    \end{scriptsize}
    \label{tab:catalogue_beispiel}
\end{sidewaystable}
%\end{landscape}

\end{appendix}

\end{document}